\documentclass[traditabstract,twocolumn]{aa}

\usepackage{graphicx}
\usepackage{xcolor,soul}
\usepackage{svg}
\usepackage{longtable}
\usepackage{array}
\usepackage{lscape}
\usepackage{adjustbox}
\usepackage{xcolor}
\usepackage[toc,page]{appendix}
\usepackage{gensymb}

\usepackage{txfonts}
\usepackage{amsmath,siunitx}

\usepackage[breaklinks, colorlinks, citecolor=blue, linkcolor=blue]{hyperref}

\usepackage{xcolor}

\usepackage{orcidlink}
\usepackage{hyperref}

\providecommand{\mj}{\ensuremath{\,{\rm M_{Jup}}}}
\providecommand{\rj}{\ensuremath{\,{\rm R_{Jup}}}}

\providecommand{\mst}{\ensuremath{\,{\rm M_\odot}}}
\providecommand{\rst}{\ensuremath{\,{\rm R_\odot}}}
\providecommand{\lst}{\ensuremath{\,{\rm L_\odot}}}

\providecommand{\arcsec}{$^{\prime \prime}$}

\graphicspath{{./}{figures/}}

\begin{document}

   \title{BD-14\,3065b (TOI-4987b): from giant planet to brown dwarf: evidence for deuterium burning in old age?
   }

   \subtitle{}

   \author{J\'an \v{S}ubjak 
          \inst{1,2}\orcidlink{0000-0002-5313-9722}
          \and
          David W. Latham
          \inst{2}\orcidlink{0000-0001-9911-7388}
          \and
          Samuel N. Quinn
          \inst{2}\orcidlink{0000-0002-8964-8377}
          \and
          Perry Berlind
          \inst{2}\orcidlink{0009-0005-7108-9502}
          \and
          Michael L. Calkins
          \inst{2}\orcidlink{0000-0002-2830-5661}
          \and
          Gilbert A. Esquerdo
          \inst{2}\orcidlink{0000-0002-9789-5474}
          \and
          Rafael Brahm
          \inst{3,4,5}\orcidlink{0000-0002-9158-7315}
          \and
          Jos\'e A. Caballero
          \inst{6}\orcidlink{0000-0002-7349-1387}
          \and
          Karen A. Collins
          \inst{2}\orcidlink{0000-0001-6588-9574}
          \and
          Eike Guenther
          \inst{7}\orcidlink{0000-0002-9130-6747}
          \and
          Jan Jan\'ik
          \inst{8}\orcidlink{0000-0002-6384-0184}
          \and
          Petr Kab\'ath
          \inst{1}\orcidlink{0000-0002-1623-5352}
          \and
          Richard P. Schwarz
          \inst{2}\orcidlink{0000-0001-8227-1020}
          \and
          Thiam-Guan Tan
          \inst{9}\orcidlink{0000-0001-5603-6895}
          \and
          Leonardo Vanzi
          \inst{10}
          \and
          Roberto Zambelli
          \inst{11}
          \and
          Carl Ziegler
          \inst{12}\orcidlink{0000-0002-0619-7639}
          \and
          Jon M. Jenkins
          \inst{13}\orcidlink{0000-0002-4715-9460}
          \and
          Ismael Mireles
          \inst{14}\orcidlink{0000-0002-4510-2268}
          \and
          Sara Seager
          \inst{15,16,17}\orcidlink{0000-0002-6892-6948}
          \and
          Avi Shporer
          \inst{15}\orcidlink{0000-0002-1836-3120}
          \and
          Stephanie Striegel
          \inst{18}\orcidlink{0009-0008-5145-0446}
          \and
          Joshua N.\ Winn
          \inst{19}\orcidlink{0000-0002-4265-047X}}

   \institute{Astronomical Institute, Czech Academy of Sciences, Fri{\v c}ova 298, 251 65, Ond\v{r}ejov, Czech Republic
         \and
         Center for Astrophysics ${\rm \mid}$ Harvard {\rm \&} Smithsonian, 60 Garden Street, Cambridge, MA 02138, USA
         \and
         Facultad de Ingeniera y Ciencias, Universidad Adolfo Ib\'{a}\~{n}ez, Av. Diagonal las Torres 2640, Pe\~{n}alol\'{e}n, Santiago, Chile
         \and
         Millennium Institute for Astrophysics, Chile
         \and
         Data Observatory Foundation, Chile
         \and
         Centro de Astrobiolog\'ia (CSIC-INTA), ESAC, Camino bajo del castillo s/n, 28692 Villanueva de la Ca\~{n}ada, Madrid, Spain
         \and
         Th\"uhringer Landessternwarte Tautenburg, Sternwarte 5, 07778 Tautenburg, Germany
         \and
         Department of Theoretical Physics and Astrophysics, Faculty of Science, Masaryk University, Kotlarska 2, CZ-611 37, Brno, Czech Republic
         \and
         Perth Exoplanet Survey Telescope (PEST), Perth, Western Australia
         \and
         Department of Electrical Engineering and Centre of Astro-Engineering, Pontificia Universidad Catolica de Chile, Av. Vicu\~{n}a Mackenna 4860, Santiago, Chile
         \and
         Societa' Astronomica Lunae, Italy
         \and
         Department of Physics, Engineering and Astronomy, Stephen F. Austin State University, 1936 North St, Nacogdoches, TX 75962, USA
         \and
         NASA Ames Research Center, Moffett Field, CA 94035, USA
         \and
         Department of Physics and Astronomy, University of New Mexico, 210 Yale Blvd NE, Albuquerque, NM 87106, USA
         \and
         Department of Physics and Kavli Institute for Astrophysics and Space Research, Massachusetts Institute of Technology, Cambridge, MA 02139, USA
         \and
         Department of Earth, Atmospheric and Planetary Sciences, Massachusetts Institute of Technology, Cambridge, MA 02139, USA
         \and
         Department of Aeronautics and Astronautics, MIT, 77 Massachusetts Avenue, Cambridge, MA 02139, USA
         \and
         SETI Institute, Mountain View, CA 94043 USA/NASA Ames Research Center, Moffett Field, CA 94035 USA
         \and
         Department of Astrophysical Sciences, Princeton University, Princeton, NJ 08544, USA
        }

   \date{\today{}; \today{}}

 
\abstract{The present study reports the confirmation of BD-14\,3065b, a transiting planet/brown dwarf in a triple-star system, with a mass near the deuterium burning boundary. BD-14\,3065b has the largest radius observed within the sample of giant planets and brown dwarfs around post-main-sequence stars. Its orbital period is 4.3 days, and it transits a subgiant F-type star with a mass of $M_\star=1.41 \pm 0.05$\,M$_{\odot}$, a radius of $R_\star=2.35 \pm 0.08$\,R$_{\odot}$, an effective temperature of $T_{\rm eff}=6935\pm90$\,K, and a metallicity of $-0.34\pm0.05$\,dex. By combining TESS photometry with high-resolution spectra acquired with the TRES and Pucheros+ spectrographs, we measured a mass of $M_p=12.37\pm0.92\,\mj$ and a radius of $R_p=1.926\pm0.094\,\rj$. Our discussion of potential processes that could be responsible for the inflated radius led us to conclude that deuterium burning is a plausible explanation resulting from the heating of BD-14\,3065b's interior. Detection of the secondary eclipse with TESS photometry enables a precise determination of the eccentricity $e_p=0.066\pm0.011$ and reveals BD-14\,3065b has a brightness temperature of $3520 \pm 130$\,K. With its unique characteristics, BD-14\,3065b presents an excellent opportunity to study its atmosphere through thermal emission spectroscopy.
}

   \keywords{giant planets -- brown dwarfs -- techniques: photometric -- techniques: spectroscopic -- techniques: radial velocities
               }
\maketitle

%
%
%

\section{Introduction}\label{sec:introduction}

The question of distinguishing between planets and brown dwarfs has been debated at length, with a proposed solution being the use of formation history as the most fundamental differentiator \citep{burrows01}. According to this approach, substellar objects formed through disk instability similar to stars are classified as brown dwarfs, while those formed via core accretion are classified as planets. However, determining the formation history of individual systems observationally is a difficult task. Some authors used the lack of known transiting brown dwarfs in the mass range between $35 \leq \mj \sin{i} \leq 55$ to motivate the idea that this gap separates two different BD populations that result from different BD formation processes \citep{Ma2014}. Recent discoveries of new transiting brown dwarfs in this region have revealed that these two formation mechanisms do not form such a well-defined mass boundary. Furthermore, it was theorized that disk instability can produce objects down to a few masses of Jupiter, while core accretion can produce objects as massive as tens of Jupiter \citep{Mordasini12}.

Another proposed way to differentiate between planets and brown dwarfs is predicated on the deuterium reactions occurring within their interior, as outlined by \cite{Saumon96}, \cite{Chabrier00}, or \cite{burrows01}. The mass of an object serves as the primary determinant in this definition, allowing for straightforward classification in most cases. However, objects that fall near the dividing line pose difficulties, particularly given the uncertainties surrounding their mass and chemical composition. Many authors have adopted the dividing line of $13\,\mj$; nevertheless, every attempt to classify objects close to this boundary is akin to shooting in the dark. As demonstrated by \cite{Spiegel11}, or \cite{Moliere12}, deuterium burning is reliant on the deuterium abundance, helium mass fraction, and metallicity, leading to a blurring of the dividing line from $11-16\,\mj$ depending on the specific chemical composition.

In their paper, \cite{Hatzes15} presented an argument for defining planets and brown dwarfs based on observables such as mass and radius. While objects in the mass range of 0.3 to 80 Jupiter masses follow a consistent trend in the mass-radius diagram, the authors contend that there is no compelling reason to classify brown dwarfs differently. An observer might argue that we should investigate whether objects above and below $13\,\mj$ are distinct. Therefore, objects that straddle this threshold, such as BD-14\,3065b, present an intriguing opportunity for further study. If these objects can undergo radius re-inflation during a post-main-sequence phase of the host star due to deuterium burning, they may indeed differ from their lower-mass counterparts.

A critical inquiry shared by transiting giant planets and brown dwarfs pertains to their inflated radii, given that many of these objects appear to have an unexpectedly large radius. Notably, the typical radii of known brown dwarf companions range from 0.7 to 1.4 Jupiter radii \citep{Subjak20,Carmichael21,Subjak23}, while some hot Jupiters are approaching a radius of 2 Jupiter radii \citep[e.g.,][]{Anderson11,Fulton15,Almenara15,Yee23}. Numerous studies have been conducted to investigate hot Jupiters, revealing a correlation between the incident stellar irradiation and the measured radii of such planets \citep[e.g.,][]{Enoch12,Sestovic18}. Theoretical studies have explored the influence of stellar irradiation on the radii of companions, which has been incorporated into various evolutionary models of giant planets and brown dwarfs \citep{Baraffe03,Fortney07,Baraffe08}. In addition to this, several other factors may also play important roles in the inflation of planet radii. These include tidal dissipation \citep{Bodenheimer01,Bodenheimer03,Arras10,Jermyn17}, kinetic heating (a fraction of the incoming stellar flux is assumed to be converted into kinetic energy and then dissipated at the center of the planet) \citep{Guillot02}, enhanced atmospheric opacities \citep{Burrows07}, double-diffusive convection \citep{Chabrier07,Kurokawa15}, vertical advection of potential temperature \citep{Youdin10,Tremblin17}, Ohmic heating through magnetohydrodynamic effects \citep{Batygin10,Perna10,Wu13,Ginzburg16}, or deuterium burning \citep{Moliere12}. The mass of giant planets and brown dwarfs is a key factor in determining the extent of inflation, with the response decreasing as mass increases.

We have confirmed BD-14\,3065b as an inflated giant planet/brown dwarf orbiting a sub-giant F-type star. Our paper includes a description of the observations in Section \ref{sec:observations}, data analysis in Section \ref{sec:analysis}, a discussion in Section \ref{sec:discussion}, and a summary of our results in Section \ref{sec:summary}.


%
%

\section{Observations}\label{sec:observations}

\subsection{TESS light curves}\label{sec:TESS}

TESS monitored BD-14\,3065 in three different sectors. Its first observation was in sector 9, from February 28, 2019, to March 26, 2019, with 1800\,s time sampling. The second observation was in sector 35, from February 09, 2021, to March 07, 2021, with 600\,s time sampling. Lastly, it was observed in sector 62, from February 12, 2023, to March 10, 2023, with 120\,s time sampling. The data for BD-14\,3065 are publicly available on the Mikulski Archive for Space Telescopes (MAST)\footnote{\url{https://mast.stsci.edu/portal/Mashup/Clients/Mast/Portal.html}} and are provided by the TESS Science Processing Operations Center (SPOC) located at NASA Ames Research Center. The transit signature of BD-14\,3065b was first detected by the MIT Quick Look Pipeline \citep[QLP;][]{Huang20,Huang2020} and alerted by the TESS Science Office on January 06, 2022 \citep{Guerrero21}.

The light curves (LCs) from sectors 9 and 35 were processed by the QLP pipeline, using the difference imaging technique. This involves subtracting each frame from a reference frame generated by combining many comparison frames. Once the difference image is obtained, the QLP performs aperture photometry and scales the difference relative to the star’s average flux. This ensures that any contamination from other stars within the same aperture is automatically accounted for. The LC from TESS sector 62 processed by the SPOC pipeline \citep{Jenkins16} was downloaded directly from the MAST archive, using the {\tt lightkurve} package \citep{Lightkurve18}. The SPOC pipeline aims to remove systematic errors related to the instrument. The Pixel Response Functions (PRFs) were used to analyze the crowding, and a crowding correction was included in the PDCSAP flux time series. Neglecting such a correction can result in underestimating the planet's radius. The pipeline shows that 0.997 of the light in the optimal aperture comes from the target rather than other stellar sources, indicating insignificant dilution due to faint background stars. However, the analysis cannot rule out other stellar sources unresolved by Gaia (see Section \ref{sec:ao_image}). Moreover, according to Data Validation's \citep{Twicken18} difference image centroiding analysis, the source of the transit signature was within $1.0\pm2.5$ arcsec of the target star.

\begin{figure*}
\centering
\includegraphics[width=1.0\textwidth,height=1.1\textwidth]{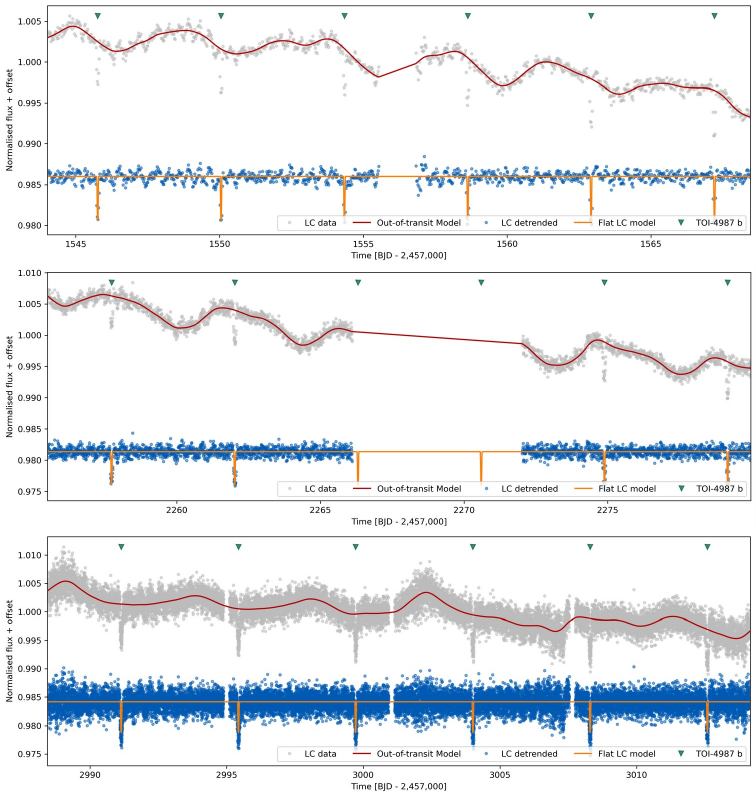}
\caption{SPOC PDCSAP LCs for BD-14\,3065 from TESS sector 9 (top panel), sector 35 (middle panel), and sector 62 (bottom panel). Grey points represent the TESS observations, while red lines correspond to the out-of-transit GP models created with {\tt citlalicue} to capture the variability in the LCs. Datasets were divided by these models leading to flattened TESS LCs (blue points) with transit models (orange lines). Green triangles indicate the positions of transits. The LCs still show pulsations, but these were later removed.}
\label{fig:tess_lc}
\end{figure*}

We used the python package called {\tt citlalicue} \citep{Barragan22} to correct for systematics and eliminate stellar variability in the PDCSAP LCs \citep{Smith12,Stumpe2012,Stumpe2014} obtained through {\tt lightkurve}. {\tt citlalicue} combines a Gaussian Process regression with transit models computed via the {\tt pytransit} code \citep{Parviainen15} to generate a model comprising both the light curve variability and the transits. The resulting model is then used to eliminate the variability and produce a flattened LC containing only the transit photometric variations. Fig. \ref{fig:tess_lc} displays the LCs before and after this process. We detected 16 transits, with six each in Sectors 9 and 62 and four in Sector 35. Additionally, the SPOC data validation report has reported the secondary eclipse signature in sector 62 with a depth of $553 \pm 183$\,ppm. We validate this signal and discuss the transit and secondary eclipse modeling in Section \ref{sec:RVs}.

\begin{figure}
\centering
\includegraphics[width=0.46\textwidth, trim= {0.0cm 0.0cm 0.0cm 0.0cm}]{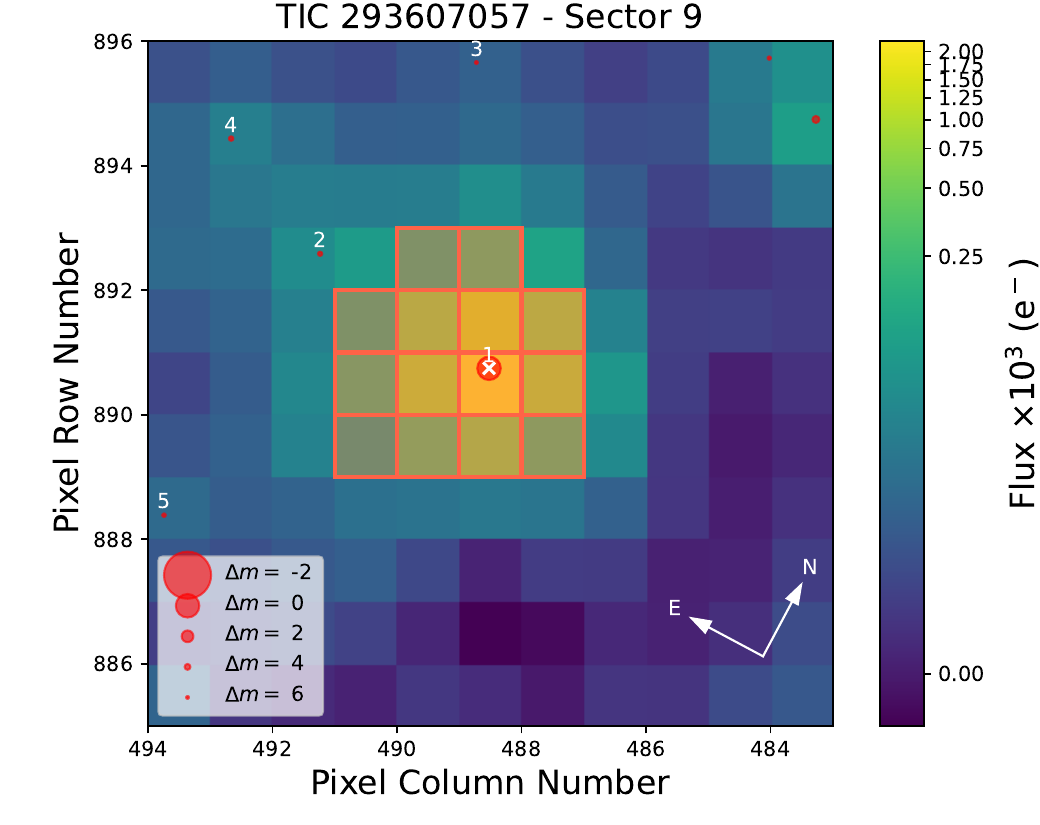}
\caption{Gaia DR2 catalog overplotted to the TESS TPF image.} \label{fig:field_image}
\end{figure}

We used a tool called {\tt tpfplotter} \citep{Aller20} to compare the Gaia DR2 catalog with the TESS target pixel file (tpf). It allowed us to detect any potential stars that could dilute the TESS photometry, with a limit of up to 6 magnitudes difference. The resulting tpf image can be viewed in Fig. \ref{fig:field_image}. We found no additional sources in the TESS aperture that could dilute TESS LCs. It is worth noting that the Gaia DR3 catalog doesn't contain any additional bright sources within the TESS aperture. Therefore, the use of its older release in our analysis is justified.

\begin{figure}
\centering
\includegraphics[width=0.46\textwidth, trim= {0.0cm 0.0cm 0.0cm 0.0cm}]{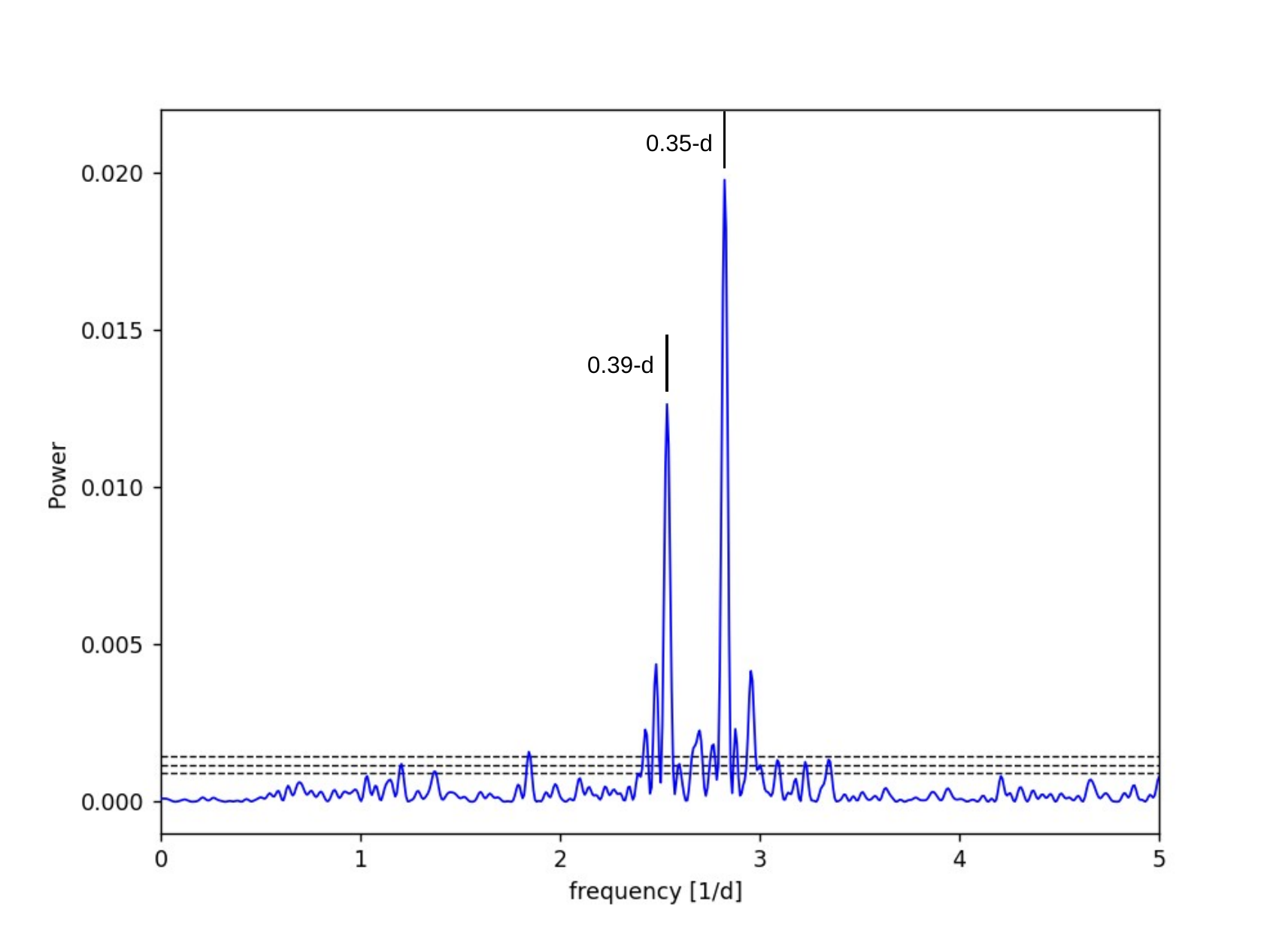}
\caption{The GLS periodogram of SPOC LCs after substruction of the 4.3-d rotational signal using Gaussian processes. Two dominant peaks represent stellar pulsations.} \label{fig:GLS_per}
\end{figure}

The SPOC LCs revealed additional stellar variability in the form of pulsations. To analyze these pulsations, we utilized maximum-likelihood (MLP) and generalized Lomb-Scargle (GLS) periodograms, which detected two signals representing pulsations with periods of 0.35 and 0.39 days, consistent across all three sectors. The GLS periodogram is shown in Figure \ref{fig:GLS_per}. We subtracted these signals by fitting two simple sinusoidal functions to the data. As a result, the final LCs appeared flat, without any signs of stellar variability.

\begin{table}
 \centering
 \caption[]{Basic parameters for BD-14\,3065.}
 \label{tab:basicpar}
 \begin{adjustbox}{width=0.48\textwidth}
    \begin{tabular}{lccr}
        \hline
	\hline
	Parameter & Description & Value & Source\\
        \hline
    $\alpha_{\rm J2000}$ & Right Ascension (RA) & 10 14 40.73 & 1\\
    $\delta_{\rm J2000}$ & Declination (Dec) & -15 38 34.33 & 1\\
    \smallskip\\
    $V_T$ & Tycho-2 $V_T$ mag & $11.146 \pm 0.098$ & 2\\
    $G$ & Gaia $G$ mag & $10.921 \pm 0.002$ & 1\\
    $T$ & TESS $T$ mag & $10.631 \pm 0.008$ & 3\\
    $J$ & 2MASS $J$ mag & $10.143 \pm 0.026$ & 4\\
    \smallskip\\
    $\mu_\alpha \cos{\delta}$ & PM in RA (mas/yr) & $-4.990 \pm 0.074$ & 1\\
    $\mu_{\rm \delta}$ & PM in DEC (mas/yr) & $7.629 \pm 0.062$ & 1\\
    $\pi$ & Parallax (mas) & $1.765 \pm 0.055$ & 1\\
    \hline
    \multicolumn{3}{l}{Other identifiers:} \\
    \multicolumn{3}{l}{TOI-4987} \\
    \multicolumn{3}{l}{TIC 293607057} \\
    \multicolumn{3}{l}{TYC 6059-01368-1} \\
    \multicolumn{3}{l}{2MASS J10144071-1538344} \\
    \multicolumn{3}{l}{Gaia DR2 3751877374435102720} \\
    \hline
    \hline
    \end{tabular}
    \end{adjustbox}
    \smallskip\\
    References: 1 - $Gaia$ DR3, \citet{Gaia21}; \\ 2 - Tycho-2, \citet{Hog00}; 3 - TESS, \citet{Stassun18}; \\ 4 - 2MASS, \citet{Skrutskie06}
\end{table}

\subsection{Contamination from nearby sources}\label{sec:ao_image}

\begin{figure}
\centering
\includegraphics[width=0.46\textwidth, trim= {0.0cm 0.0cm 0.0cm 0.0cm}]{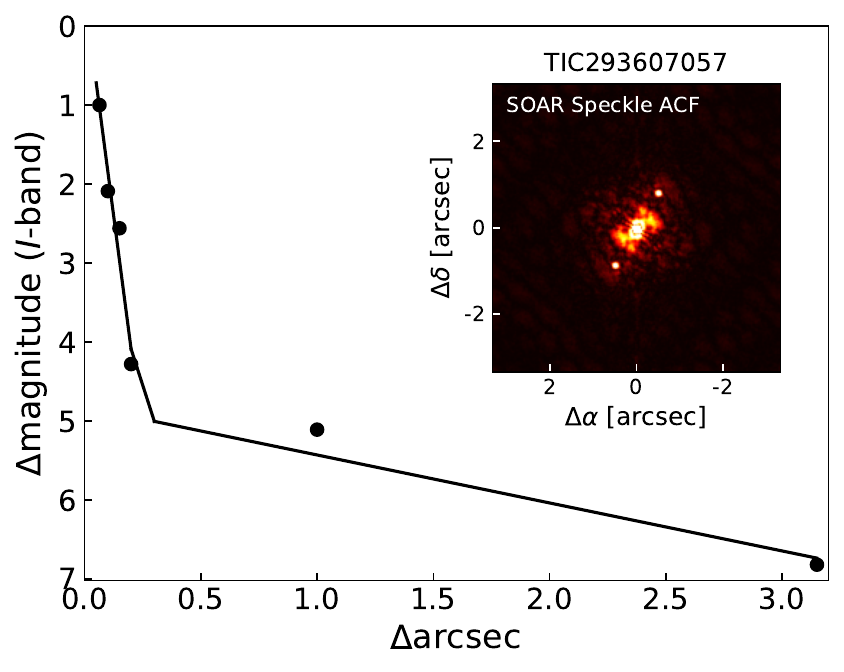}
\caption{High-resolution imaging data for BD-14\,3065 in the $I$ band. SOAR HRCam speckle sensitivity curve (solid line) and autocorrelation function (ACF, an inside image). The detected companion appears as two dots in the ACF image.} \label{fig:speckle_image}
\end{figure}

As part of follow-up observations coordinated by the TESS Follow-up Observing Program (TFOP) High-Resolution Imaging Sub-Group 3 (SG3), we obtained high angular-resolution imaging of BD-14\,3065. Observations were made on April 15, 2022, using the High-Resolution Camera \citep[HRCam;][]{Tokovinin08} speckle interferometry instrument on the Southern Astrophysical Research (SOAR) 4.1m telescope. The observation strategy and data reduction procedures are described in \cite{Tokovinin18}, or \cite{Ziegler21}. Observations were made in the Cousins $I$ filter with a resolution of 36\,mas. The final reconstructed image, shown in Fig. \ref{fig:speckle_image}, reaches a contrast of $\Delta$$mag=5.1$ at a separation of 0.5\,arcsec and has an estimated PSF that is 0.064\,arcsec wide. The speckle imaging of BD-14\,3065 detected a companion at an angular separation of 0.92\,arcsec from the primary (translated into a projected physical separation of 520\,au), which has $\Delta$$I=2.3$\,mag, and position angle of 210.5 or 30.5 degrees. In order to conduct further analyses, we adopted a conservative errorbar of 0.1 mag for magnitude difference. We also note that Gaia did not report this companion. However, the Gaia Renormalised Unit Weight Error (RUWE) value of 3.5 for BD-14\,3065 suggests the presence of a close multiplicity. We discuss the treatment of the companion in Section \ref{st_par}.

\subsection{Ground-based light curves}

As part of the TFOP, ground-based photometric data were collected for BD-14\,3065. The observations were scheduled using Transit Finder, a customized version of the {\tt Tapir software} \citep{Jensen13}, and the photometric data was extracted using {\tt AstroImageJ} \citep{Collins17}. 

Our team conducted four transits using the Las Cumbres Observatory Global Telescope \citep[LCOGT;][]{Brown13} 1.0-m network. The telescopes are equipped with 4096×4096 pixel SINISTRO cameras that have an image scale of 0.389\,arcsec per pixel. Two transits were observed in alternating filter mode, resulting in a total of four light curves. Together, three transits are in the Sloan $g'$ filter, and three transits in the Sloan $i'$ filter. The images were calibrated using the standard {\tt LCOGT BANZAI} pipeline \citep{McCully18}. The aperture radius used ranges from 5.5 to 10.5\,arcsec.

Additionally, one transit was observed with the 0.305m Perth Exoplanet Survey Telescope (PEST) in the Rc filter. The telescope is equipped with a QHY183M camera with an image scale of 0.71\,arcsec per pixel. The data reduction and aperture photometry were performed using a custom software package PEST Pipeline\footnote{\url{https://pestobservatory.com/the-pest-pipeline/}}. The aperture radius used for the observations was 5.7 \,arcsec.

\subsection{TRES spectra}
We obtained spectra of BD-14\,3065 using the Tillinghast Reflector Echelle Spectrograph (TRES) on the 1.5-m telescope at the Whipple Observatory on Mount Hopkins, Arizona, from January 22, 2022 to April 13, 2024. The spectrograph has a resolving power of R\,$\approx$\,44\,000 and covers wavelengths ranging from 390\,nm to 910\,nm. A total of 57 spectra of BD-14\,3065 were captured with TRES, with exposure times varying from 300--3600\,s depending on the weather conditions. The S/N ratio per resolution element of these spectra ranged from 24 to 60. We derived the relative RVs from the TRES spectra by considering multiple echelle orders from each spectrum, which were cross-correlated with a high SNR median spectrum produced using the complete set of observed spectra. We have also adjusted the spectra to account for the small drifts in the TRES zero point over the period of the observations. The TRES fiber diameter measures 2.4 arcseconds, and contamination from the wide stellar companion around BD-14 3065A, as resolved by speckle imaging, may be expected. The exact level of contamination also depends on the seeing. The final list of radial velocities (RVs) can be found in Table \ref{tab:long}.

\subsection{Spectroscopy from the PLATOSpec project}

We acquired additional spectra from the PLATOSpec project using the fully operational telescope E152 at La Silla, Chile. An interim spectrograph, PUCHEROS+, has been installed for scientific use. PUCHEROS+ is an improved version of a PUCHEROS spectrograph \citep{10.1111/j.1365-2966.2012.21382.x}, with a resolving power of R\,$\approx 18,000$. The Andor iKon M detector provides a wavelength range of over 400 to 700\,nm. Typically, the spectrograph's radial velocity stability is about 100 m\,s$^{-1}$ over a month for cool main sequence stars with V$=8$\,mag. PUCHEROS+ uses the ceres+ pipeline \citep{2017PASP..129c4002B}, which produces wavelength calibrated and optimally extracted order-by-order spectra. The radial velocities are obtained using the cross-correlation method from the $\mathrm{ceres+}$ pipeline. The PUCHEROS+ fiber diameter has 2.0 arcseconds, and expected contamination from the wide stellar companion around BD-14 3065A is much smaller than for the TRES spectra. While the data will be available in the ESO archive later, they are currently accessible upon request from the PLATOSpec team. A total of 34 spectra of BD-14\,3065 were captured between May 2023 and March 2024, with each exposure lasting 1800 seconds.

%
%

\section{Analysis} \label{sec:analysis}

\subsection{Modeling Stellar Parameters} \label{st_par}
We analyzed the spectra of BD-14\,3065 to determine the spectral parameters of the star. Then, we used the catalog photometry from Gaia, 2MASS, Tycho-2, Galex, and WISE, along with the $\Delta$$mag$ measured by SOAR HRCam, to model blended two-component and three-component models of the SED using the {\tt isochrones} package \citep{Morton15,Dotter16,Choi16}. The two-component model accounts for the stellar companion resolved in the speckle imaging, while the three-component model additionally accounts for the long-term signal detected in the radial velocity observations. Further elaboration on this matter is presented below. We assumed that the stars are physically associated and have the same parallax as measured for the primary by Gaia, as is the case for most stars separated by $\leq1\arcsec$ \citep{Horch14,Matson18}.

\begin{table}
 \centering
 \caption{Properties of the primary star and the stellar companions. The upper panel describes the results from the two-component model, with the final column listing the observed photometry. The lower panel describes the results from the three-component model.} 
 \label{tab:compare}
 \begin{adjustbox}{width=0.48\textwidth}
    \begin{tabular}{lccccccr}
    \hline
    \hline
    Parameter & BD-14\,3065\,A & BD-14\,3065\,B & BD-14\,3065\,AB \\
    \hline
    $M_\star$ ($\rm \mst$) & $1.41 \pm 0.05$ &  $0.99 \pm 0.03$ & -\\
    $R_\star$ ($\rm \rst$)&  $2.35 \pm 0.08$ &  $0.96 \pm 0.03$ & -\\
    $\log{g}$ &  $3.85 \pm 0.03$ &  $ 4.47 \pm 0.02$ & -\\
    $T_{\rm eff}$ (K) &  $6935 \pm 90$ &  $ 6190 \pm 150$ & -\\
    $[{\rm Fe/H}]$ & $-0.34 \pm 0.05$ &  $ -0.37 \pm 0.05$ & -\\
    $v_{\rm rot} \sin{i_\star} $ (km\,s$^{-1}$) &  $22 \pm 3$ & - & -\\
    $P_{\rm rot}$ (days) & $ 4.3 \pm 0.2 $ & - & -\\
    log L ($\lst$) & $1.06 \pm 0.03$ & $0.09 \pm 0.05$ & -\\
    log Age (Gyr) & $9.35 \pm 0.05$ & $9.35 \pm 0.05$ & -\\
    $A_V$ (mag) & $0.19 \pm 0.01$ & $0.19 \pm 0.01$ & -\\
    Galex $FUV$ mag & $18.62 \pm 0.06$ & $23.67 \pm 0.53$ & $18.69 \pm 0.09$\\
    Galex $NUV$ mag & $14.49 \pm 0.03$ & $17.70 \pm 0.23$ & $14.49 \pm 0.01$\\
    Tycho-2 $B_T$ mag & $11.51 \pm 0.02$ & $14.19 \pm 0.12$ & $11.56 \pm 0.09$\\
    Gaia $BP$ mag & $11.22 \pm 0.02$ & $13.73 \pm 0.10$ & $11.046 \pm 0.003$\\
    Tycho-2 $V_T$ mag & $11.13 \pm 0.02$ & $13.62 \pm 0.10$ & $11.15 \pm 0.10$\\
    Gaia $G$ mag & $11.02 \pm 0.02$ & $13.44 \pm 0.10$ & $10.921 \pm 0.003$\\
    Gaia $RP$ mag & $10.67 \pm 0.02$ & $12.98 \pm 0.09$ & $10.510 \pm 0.004$\\
    TESS $T$ mag & $10.66 \pm 0.02$ & $12.97 \pm 0.09$ & $10.631 \pm 0.008$\\
    2MASS $J$ mag & $10.30 \pm 0.02$ & $12.47 \pm 0.08$ & $10.14 \pm 0.03$\\
    2MASS $H$ mag & $10.10 \pm 0.02$ & $12.17 \pm 0.07$ & $9.98 \pm 0.03$\\
    2MASS $K_S$ mag & $10.07 \pm 0.02$ & $12.13 \pm 0.07$ & $9.93 \pm 0.03$\\
    WISE1 mag & $10.05 \pm 0.02$ & $12.11 \pm 0.07$ & $9.84 \pm 0.02$\\
    WISE2 mag & $10.05 \pm 0.02$ & $12.11 \pm 0.07$ & $9.87 \pm 0.02$\\
    WISE3 mag & $10.03 \pm 0.02$ & $12.09 \pm 0.07$ & $9.85 \pm 0.05$\\
    \hline
    Parameter & BD-14\,3065\,A & BD-14\,3065c & BD-14\,3065\,B \\
    \hline
    $M_\star$ ($\rm \mst$) & $1.40 \pm 0.06$ &  $1.02 \pm 0.03$ & $0.98 \pm 0.02$\\
    $R_\star$ ($\rm \rst$)&  $2.18 \pm 0.08$ &  $1.02 \pm 0.05$ & $0.96 \pm 0.03$\\
    $\log{g}$ &  $3.91 \pm 0.03$ &  $4.43 \pm 0.03$ & $4.47 \pm 0.02$\\
    $T_{\rm eff}$ (K) &  $6985 \pm 90$ &  $6300 \pm 150$ & $6175 \pm 120$\\
    $[{\rm Fe/H}]$ & $-0.34 \pm 0.04$ &  $ -0.39 \pm 0.05$ & $ -0.37 \pm 0.04$\\
    $v_{\rm rot} \sin{i_\star} $ (km\,s$^{-1}$) &  $22 \pm 3$ & - & -\\
    $P_{\rm rot}$ (days) & $ 4.3 \pm 0.2 $ & - & -\\
    log L ($\lst$) & $1.00 \pm 0.03$ & $0.16 \pm 0.07$ & $0.08 \pm 0.05$\\
    log Age (Gyr) & $9.36 \pm 0.06$ & $9.36 \pm 0.06$ & $9.36 \pm 0.06$\\
    $A_V$ (mag) & $0.19 \pm 0.01$ & $0.19 \pm 0.01$ & $0.19 \pm 0.01$\\
    Galex $FUV$ mag & $18.63 \pm 0.07$ & $22.92 \pm 0.67$ & $23.77 \pm 0.49$\\
    Galex $NUV$ mag & $14.57 \pm 0.04$ & $17.33 \pm 0.33$ & $17.75 \pm 0.22$\\
    Tycho-2 $B_T$ mag & $11.66 \pm 0.04$ & $13.97 \pm 0.20$ & $14.22 \pm 0.12$\\
    Gaia $BP$ mag & $11.37 \pm 0.04$ & $13.55 \pm 0.18$ & $13.76 \pm 0.10$\\
    Tycho-2 $V_T$ mag & $11.28 \pm 0.04$ & $13.44 \pm 0.17$ & $13.65 \pm 0.10$\\
    Gaia $G$ mag & $11.17 \pm 0.04$ & $13.27 \pm 0.16$ & $13.47 \pm 0.10$\\
    Gaia $RP$ mag & $10.83 \pm 0.04$ & $12.83 \pm 0.15$ & $13.01 \pm 0.09$\\
    TESS $T$ mag & $10.82 \pm 0.04$ & $12.82 \pm 0.15$ & $13.01 \pm 0.09$\\
    2MASS $J$ mag & $10.47 \pm 0.04$ & $12.34 \pm 0.14$ & $12.50 \pm 0.08$\\
    2MASS $H$ mag & $10.28 \pm 0.04$ & $12.05 \pm 0.13$ & $12.20 \pm 0.07$\\
    2MASS $K_S$ mag & $10.25 \pm 0.04$ & $12.02 \pm 0.12$ & $12.16 \pm 0.07$\\
    WISE1 mag & $10.23 \pm 0.04$ & $12.00 \pm 0.12$ & $12.14 \pm 0.07$\\
    WISE2 mag & $10.23 \pm 0.04$ & $11.99 \pm 0.12$ & $12.14 \pm 0.07$\\
    WISE3 mag & $10.21 \pm 0.04$ & $11.97 \pm 0.12$ & $12.12 \pm 0.07$\\
    \hline\hline
    \end{tabular}
 \end{adjustbox}
\end{table}

\begin{figure*}
\centering
\includegraphics[width=1.0\textwidth,height=0.5\textwidth]{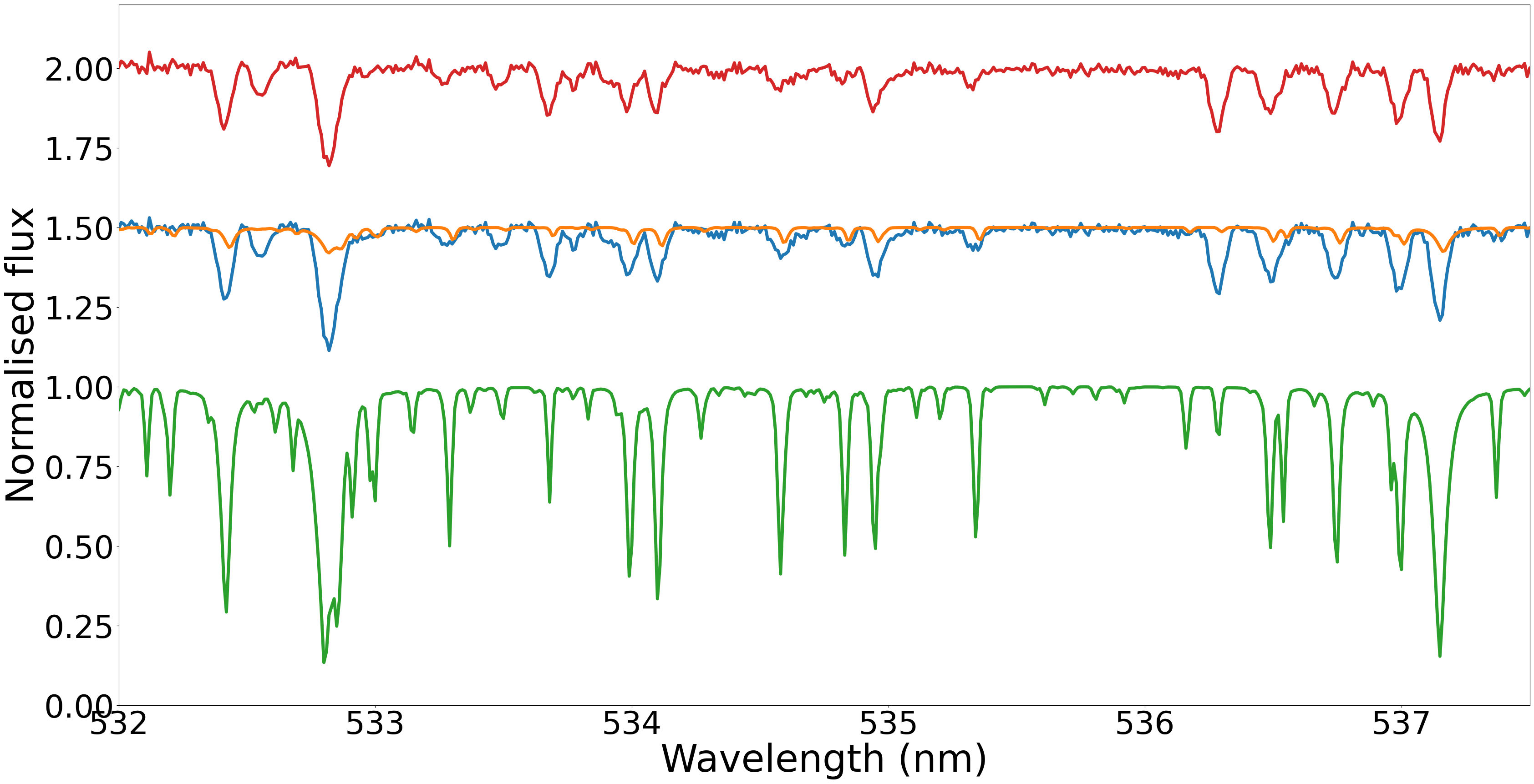}
\caption{Comparison of the observed spectrum (blue line) with the G-type star synthetic spectrum (green line) shifted by offset. The orange spectrum represents the scaled green spectrum with respect to the brightness of the blended stellar companion. The red spectrum represents the spectrum of the primary star.}
\label{fig:spectra_FG}
\end{figure*}

\subsubsection{{\tt iSpec} and {\tt isochrones} Stellar Parameters} \label{iSpec}

To obtain the spectral parameters of the host stars, we used the {\tt iSpec} framework \citep{Blanco14,Blanco19}, which incorporates the SME radiative transfer code and the MARCS atmosphere models, together with version 5 of the GES atomic line list. We co-added all TRES spectra to obtain the high S/N final spectrum. This spectrum provided a higher S/N ratio than the PUCHEROS+ spectrum and was therefore used to determine stellar parameters, including the effective temperature $T_{\rm eff}$, metallicity [Fe/H], surface gravity $\log{g}$, and the projected stellar equatorial velocity $v \sin{i}$. We used a spectral fitting technique that minimized the chi-squared value between the calculated synthetic and observed spectrum in the interval from 490 to 560\,nm.

The broad-band catalog and time-series photometry of BD-14\,3065 need to be corrected for the contamination of the second star found in the speckle imaging. Hence, we used the python package called {\tt isochrones} to create a two-component model of the SED, taking into account the priors set on the derived spectral parameters, available catalog photometry, $\Delta$$I$\,mag difference from speckle imaging, and Gaia parallax. We also considered the line-of-sight extinction by placing an upper limit derived from the Bayestar dust maps of \cite{Green19} incorporated into the {\tt dustmaps} code \citep{Green18}. We have obtained the stellar parameters of both stars reported in Table \ref{tab:compare}. The spectral types used to differentiate the stars throughout the text are based on the empirical spectral type-color sequence from \cite{Pecaut13}.

\begin{figure}
\centering
\includegraphics[width=0.46\textwidth, trim= {0.0cm 0.0cm 0.0cm 0.0cm}]{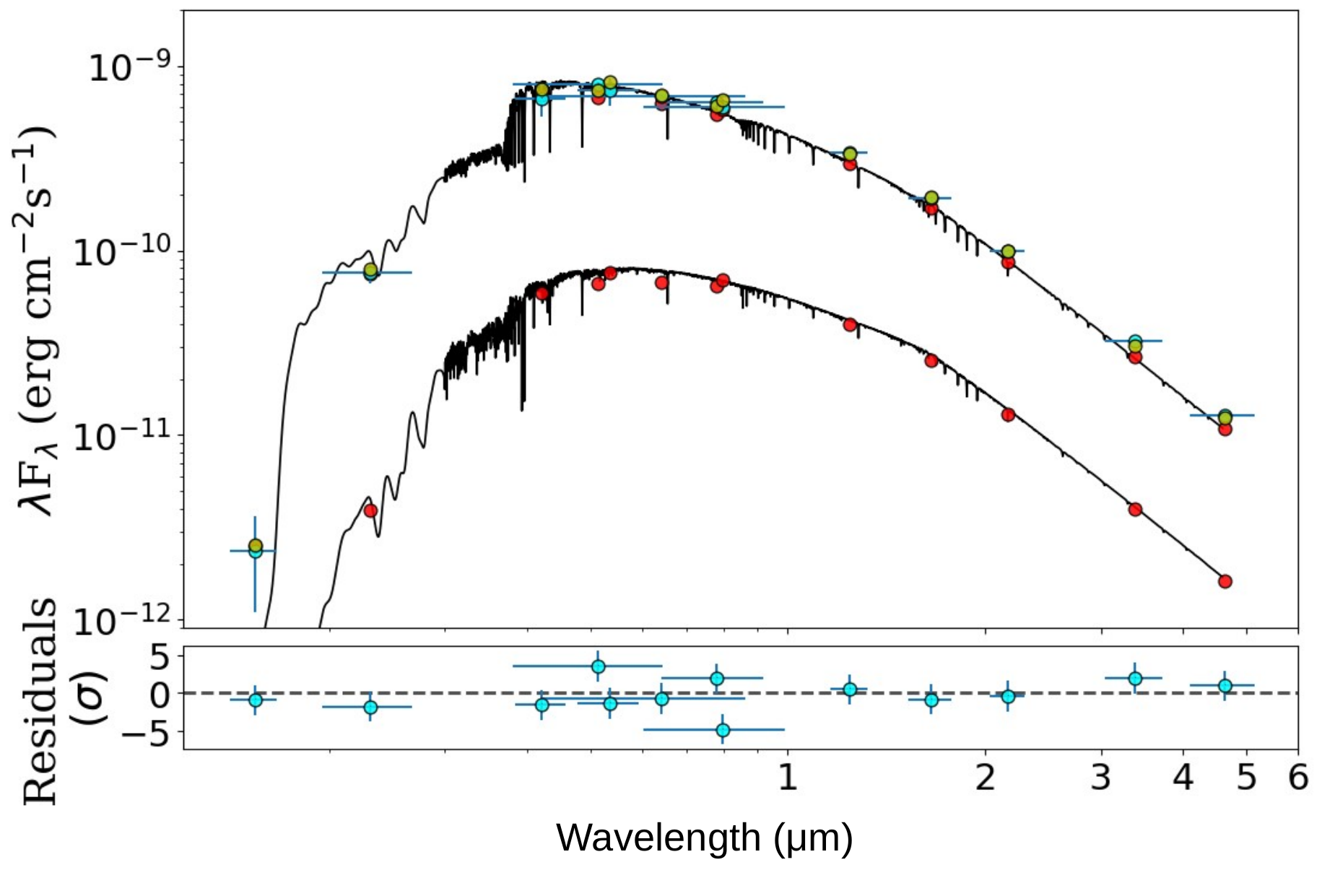}
\caption{Two-component SED fit for the BD-14\,3065 system. The blue points with vertical error bars show the observed catalog fluxes and uncertainties from Galex, Gaia, TESS, Tycho-2, 2MASS, and WISE, while the horizontal errorbars illustrate the width of the photometric band. The red points show the model fluxes of the primary and secondary, while the yellow points show the combined flux of both stars. For illustrative purposes, we overplot extinction-corrected Phoenix atmospheric models \citep{Husser2013} for the two stellar components, although these models were not used directly in the fit, which was based on MIST bolometric correction tables.} \label{fig:SED}
\end{figure}

\begin{figure}
\centering
\includegraphics[width=0.46\textwidth, trim= {0.0cm 0.0cm 0.0cm 0.0cm}]{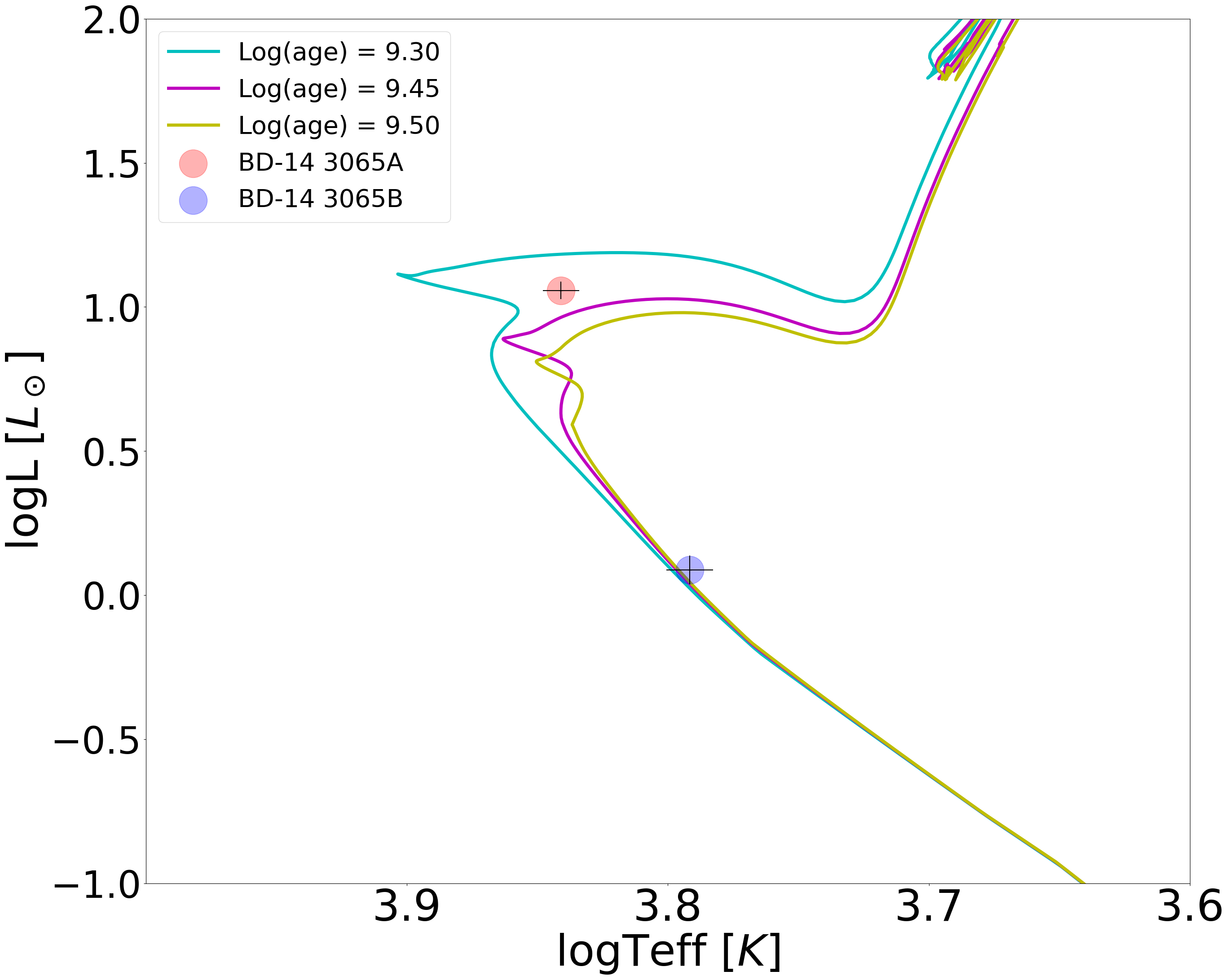}
\caption{Luminosity vs. effective temperature plot for the two-component model. Curves represent MIST isochrones for ages: 2.0\,Gyr (cyan), 2.8\,Gyr (purple), 3.2\,Gyr (yellow), and for [Fe/H]\,=\,-0.5. Red and blue points represent the positions of BD-14\,3065\,A and BD-14\,3065\,B with their errorbars.} \label{fig:isochrones}
\end{figure}

We performed a test to determine if the G-type companion star, discovered through speckle imaging, impacts our spectral parameters. To do this, we synthesized the spectrum of a G-type star with a $\Delta$$I$ of 2.3\,mag. In the wavelength range of $490-560$\,nm, the $\Delta$mag is 2.5, indicating that the companion star is ten times fainter. However, since a G-type star has deeper spectral lines compared to the F-type primary star, the companion star can still affect our findings. We also considered that a G-type star has a slower rotation rate. Figure \ref{fig:spectra_FG} shows the spectra of both stars. Our findings revealed that the contamination could cause an underestimation of the effective temperature by 200\,K and an overestimation of the value of $\log{g}$ by 0.1\,dex. It is consistent with the work of \cite{ElBadry18}, who studied the implications of unresolved binaries in stellar spectra for spectral fitting. They presented a figure that describes the effects of unresolved binarity on the accuracy of stellar labels recovered from spectral fitting as a function of the mass ratio between primary and secondary. Therefore, we adjusted our priors placed on the derived spectral parameters in SED modeling to take this effect into account. We plot the best-fit SED model in the Fig. \ref{fig:SED}, and present the stellar properties of the primary and secondary components in Table \ref{tab:compare}. We also overplot in Fig. \ref{fig:isochrones} luminosities and effective temperatures of both stars with the MIST stellar evolutionary tracks \citep{Choi16}. According to the analysis of the stellar isochrones, the system appears to have an age of approximately $2.25$\,Gyr. In order to corroborate this finding, we searched for the lithium $670.8$\,nm feature in the co-added TRES spectrum, yet did not detect any indication of it, further supporting the system's old age. Additionally, the synchronized rotation period of the star, found in Section \ref{rotation}, with the planet's orbital period suggests tidal evolution, which also aligns with the old age of the system. Finally, the {\tt PyAstronomy} python package was utilized to determine the $U$,$V$,$W$ velocities for BD-14\,3065, using the {\tt gal\_uvw}\footnote{\url{https://pyastronomy.readthedocs.io/en/latest/pyaslDoc/aslDoc/gal_uvw.html}} function. Based on our analysis, we found that BD-14\,3065 is associated with the thin disk \citep[e.g.,][]{Leggett92,Bensby03}. This outcome is predictable since the thick disk is mainly composed of stars that are at least 8\,Gyr old \citep[e.g.,][]{Fuhrmann98}.

\begin{figure}
\centering
\includegraphics[width=0.46\textwidth, trim= {0.0cm 0.0cm 0.0cm 0.0cm}]{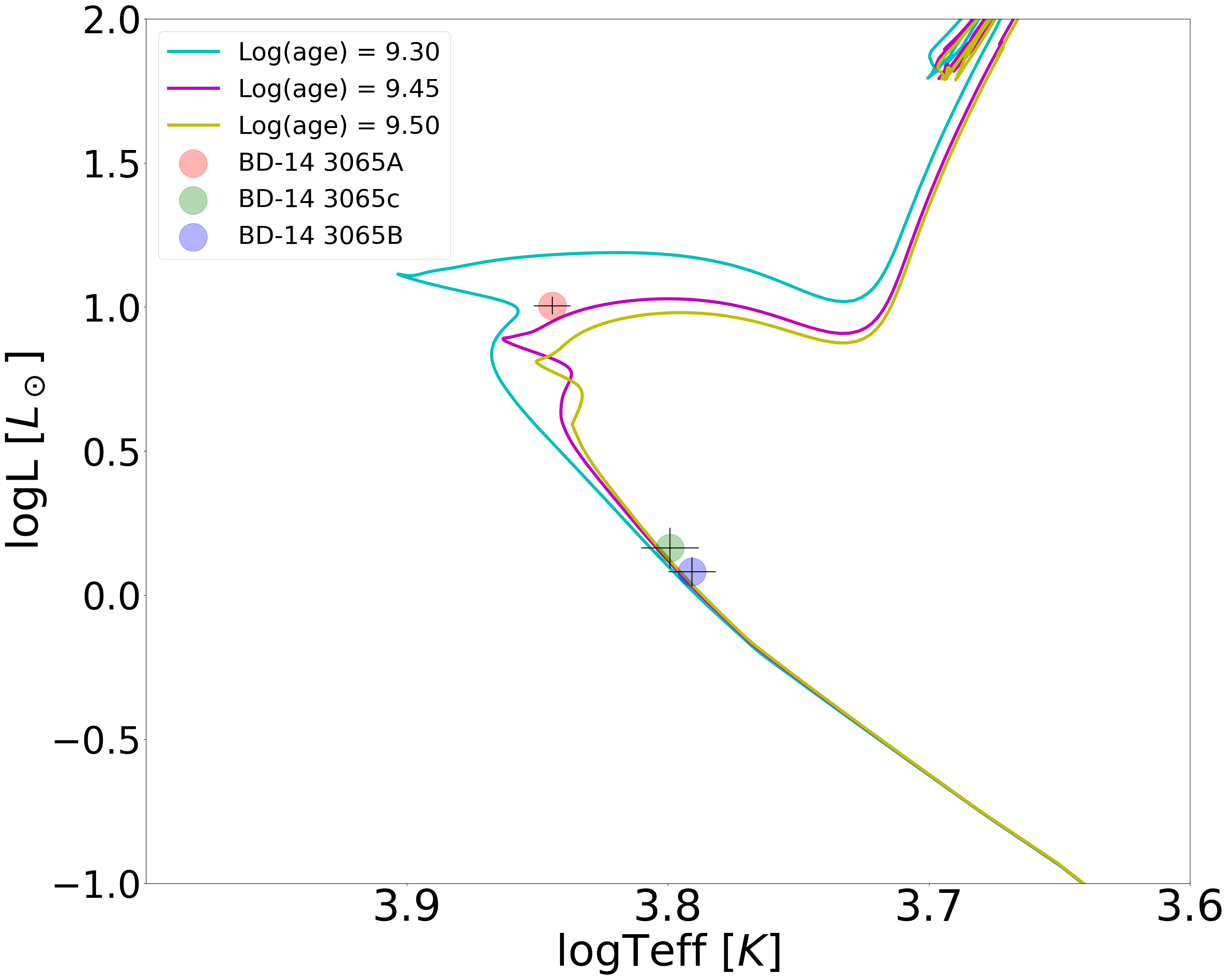}
\caption{Luminosity vs. effective temperature plot for the three-component model. Curves represent MIST isochrones for ages: 2.0\,Gyr (cyan), 2.8\,Gyr (purple), 3.2\,Gyr (yellow), and for [Fe/H]\,=\,-0.5. Red, green and blue points represent the positions of all three stars in the system with their errorbars.} \label{fig:isochrones1}
\end{figure}

In Section \ref{sec:Freq} we show an observed long-period trend in radial velocities that could be attributed to the presence of an additional star even closer than 0.9\,arcsec. We can rule out that the trend is linear; however, we cannot determine its period as it remains unconstrained. There are two possible scenarios: the star may be too faint to cause significant dilution, or it could be more massive and blended in catalog photometry. To account for the second option, we modeled a three-component spectral energy distribution with the primary star being an unresolved binary. Similarly to the two-component SED model, the input parameters for this model included the derived spectral parameters of the primary, available catalog photometry, $\Delta$$I$\,mag difference from speckle imaging, and Gaia parallax. However, in this case, the $\Delta$\,mag difference is between the star located at 0.9\,arcsec and the blended binary star at a smaller separation. The best-fit SED model is shown in Figure \ref{fig:SED1}, along with the stellar properties of all three components in Table \ref{tab:compare}. To provide further context, we also overplot stars' luminosities and effective temperatures with the MIST stellar evolutionary tracks in Figure \ref{fig:isochrones1}.

\subsubsection{Stellar rotation} \label{rotation}

The rotation period of the star was established using the SPOC TESS PDCSAP LCs. It is important to note that the long trends (see Fig. \ref{fig:tess_lc}) and transits have been removed prior to analysis. To achieve this, we made use of the Gaussian Process (GP) Regression library, {\tt Celerite}. In \cite{Foreman17}, the authors provide a detailed explanation of the library, including the interpretation of various kernels. To account for the variations in LCs due to heterogeneous surface features such as spots and plages, we utilized a rotational kernel function defined as:

\begin{figure}
\centering
\includegraphics[width=0.46\textwidth, trim= {0.0cm 0.0cm 0.0cm 0.0cm}]{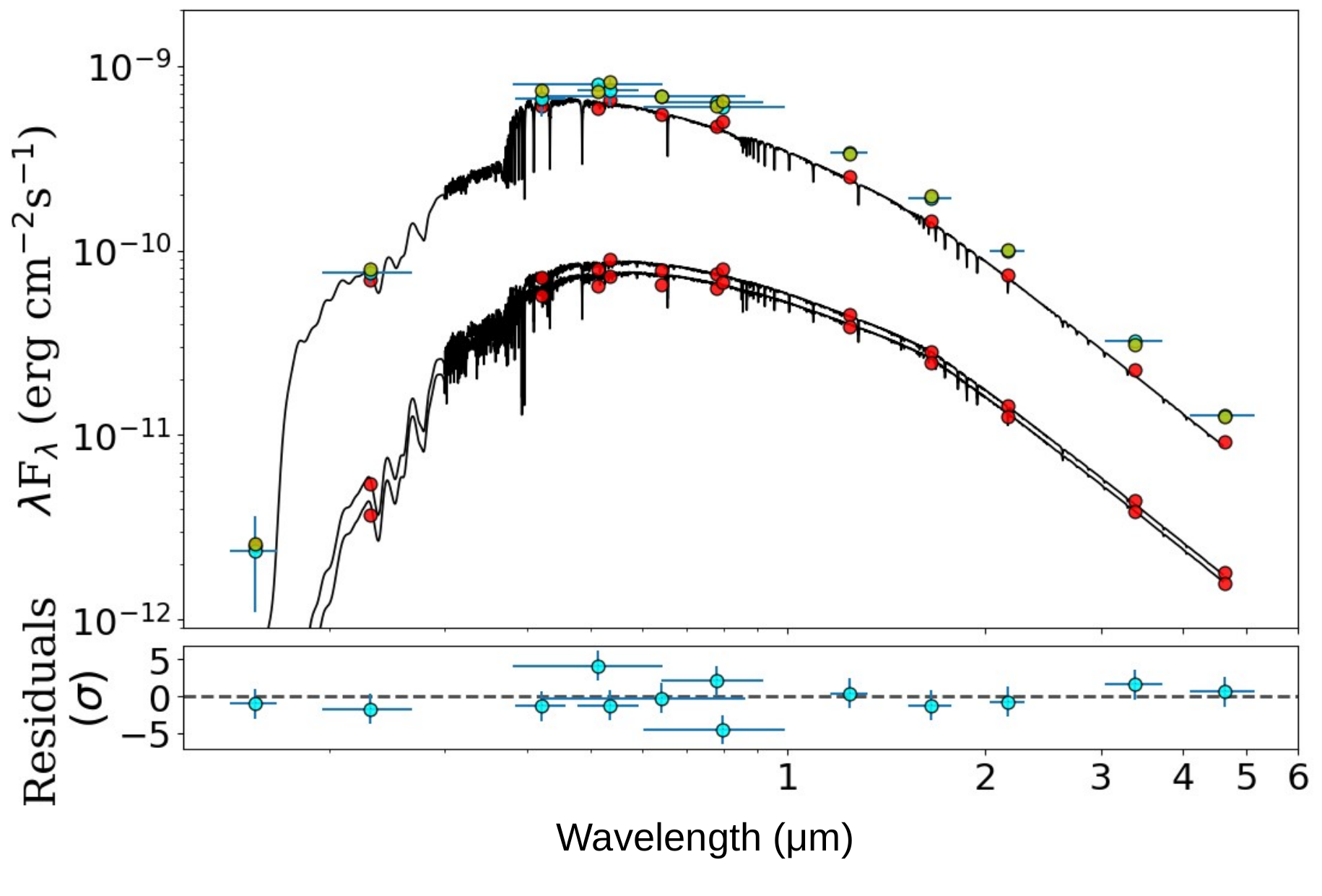}
\caption{Three-component SED fit for the BD-14\,3065 system. The blue points with vertical error bars show the observed catalog fluxes and uncertainties from Galex, Gaia, TESS, Tycho-2, 2MASS, and WISE, while the horizontal errorbars illustrate the width of the photometric band. The red points show the model fluxes of the three components, while the yellow points show the combined flux of three stars. For illustrative purposes, we overplot extinction-corrected Phoenix atmospheric models for the three stellar components, although these models were not used directly in the fit, which was based on MIST bolometric correction tables.} \label{fig:SED1}
\end{figure}

\begin{figure*}
\centering
\includegraphics[width=1.0\textwidth]{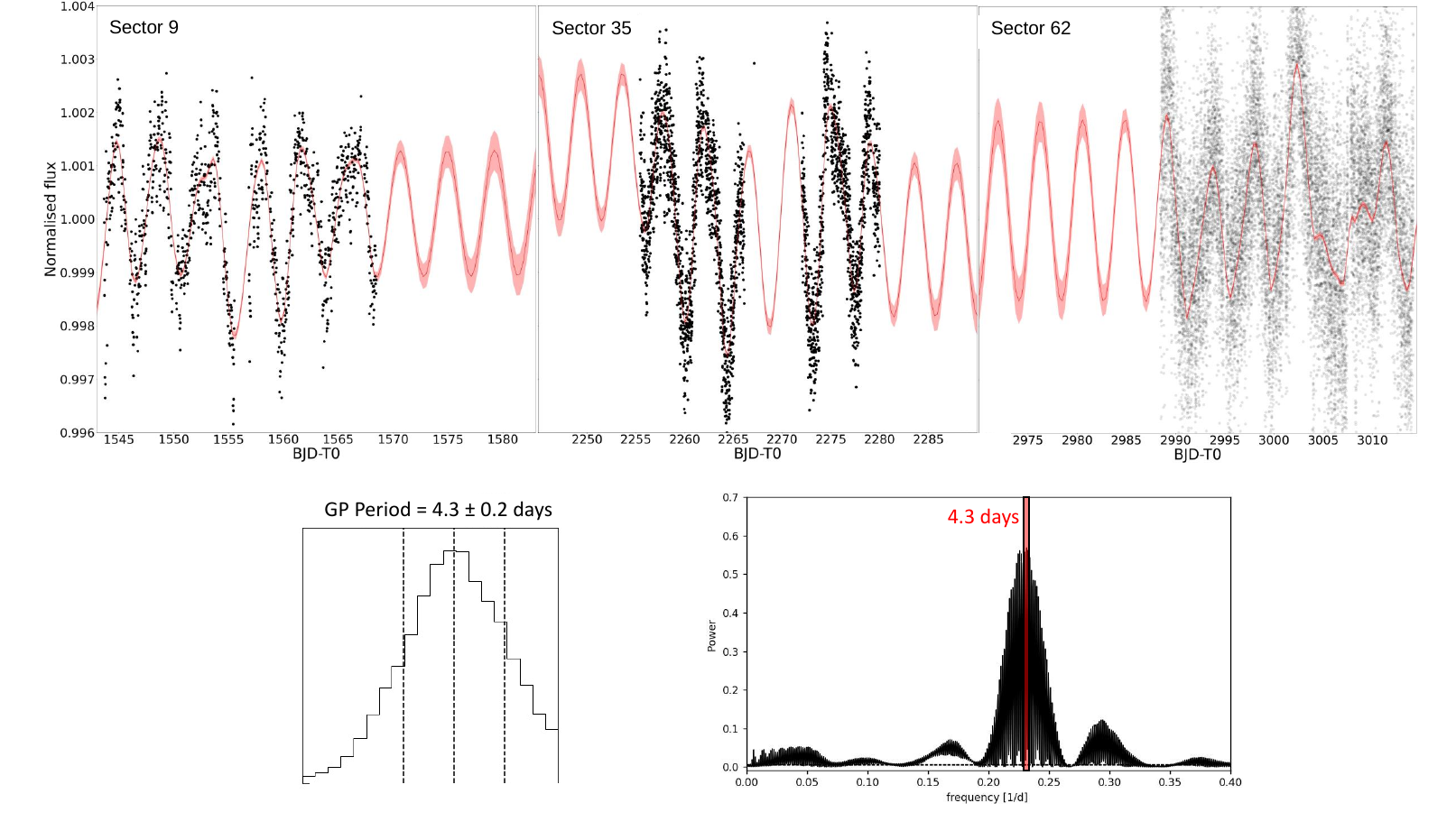}
\caption{Top: TESS SPOC PDCSAP LCs (black points) with the MAP model prediction. It is important to note that the long trends and transits have been removed prior to analysis. The red line shows the predictive mean, and the red contours show the predictive standard deviation. Bottom left: probability density of the rotation period. The period is the parameter $P_{rot}$ in Equation \ref{eq_rot}. The 1$\sigma$ error bar is indicated with the dashed black lines. Bottom right: GLS periodogram for all three sectors combined.} \label{fig:gp_plots}
\end{figure*}
\begin{figure}
\centering
\includegraphics[width=0.46\textwidth, trim= {0.0cm 0.0cm 0.0cm 0.0cm}]{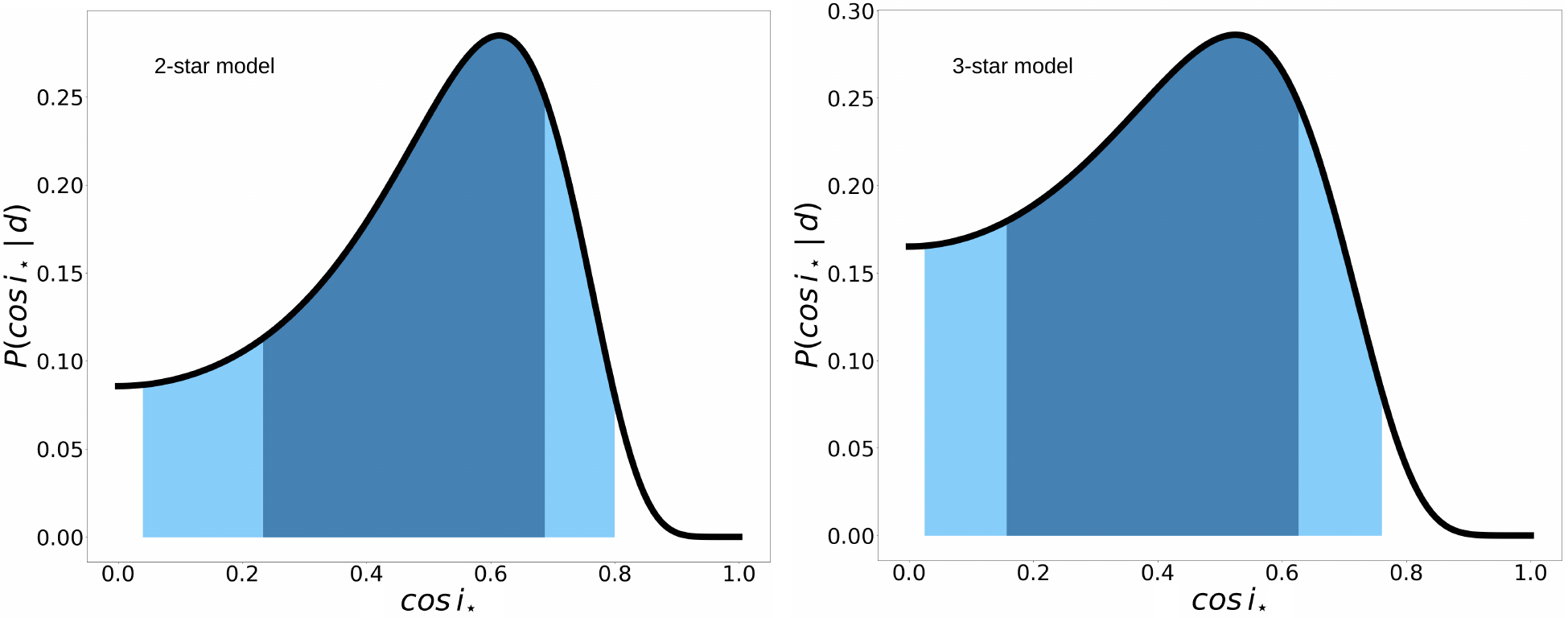}
\caption{Posterior distributions for line-of-sight stellar inclination in the form of cos\,i$_{\star}$ for the two-star (left) and three-star (right) models. Blue shadow regions show 1-$\sigma$ and 2-$\sigma$ credible intervals.} \label{fig:inclination}
\end{figure}


\begin{equation}\label{eq_rot}
      k(\tau) =
         { \frac{A}{2 + B} e^{-\tau/L} \left[ cos\left(\frac{ 2\pi\tau }{ P_{rot} }\right) + (1+B)
                   \right]
}\,,
   \end{equation}

\noindent
where variables $A$ and $B$ define the amplitude of the GP, while $\tau$ represents the time-lag and $L$ denotes the amplitude-modulation timescale of the GP. $P_{rot}$ indicates the rotational period. The maximum a posteriori (MAP) parameters are estimated using the L-BFGS-B nonlinear optimization routine \citep{Byrd95,Zhu97}. To obtain the marginalized posterior distributions of free parameters, we run MCMC using {\tt emcee} \citep{Goodman10,Foreman13}. The analysis results in a derived rotational period of $P_{rot}=4.3 \pm 0.2$\,days, which is shown in Fig. \ref{fig:gp_plots} as the probability density of $P_{rot}$ alongside the MAP model prediction. Additionally, we apply the generalized Lomb-Scargle (GLS) periodograms \citep{Zechmeister09} to the TESS LCs as an independent check of the derived $P_{rot}$. We observe strong peaks around 4.3 days in individual sectors and a forest of peaks around 4.3 days when using all sectors together. We consider this to be the rotation period, and the periodogram is shown in Fig. \ref{fig:gp_plots}. Table \ref{tab:compare} reports the derived rotation period. By using combination of $P_{rot}$, $v \sin{i}$, and $R_\star$ we can constrain stellar inclination angle, which is $i_\star\,=\,59_{-12}^{+17}$\,degrees for the two-star model and $i_\star\,=\,65_{-14}^{+16}$\,degrees for the three-star model. Our approach follows the procedure described in \cite{Masuda20} and used in \cite{Bowler23}. This method properly accounts for the correlation between stellar equatorial velocity and projected rotational velocity. Posterior distributions of stellar inclination for both models can be seen in Fig. \ref{fig:inclination}. It should be noted that the used value of $v \sin{i}$ represents an upper limit, as the spectral contamination of BD-14\,1065 can lead to an overestimation of the $v \sin{i}$ value by a few km\,s$^{-1}$. Correction for this overestimation would lead to an even higher probability of spin-orbit misalignment.

\subsection{System architecture}

In order to model the parameters of the transiting object, we need to distinguish if it orbits the primary F-type star or, rather, the wide G-type companion. A critical study that shows the significance of such analyses is presented in the work of \cite{Mandushev05}. The GSC01944-02289 system bears a strong resemblance to the architecture of the BD-14 3065 system, with an F-type primary star and a wide G-type companion. The research revealed that the potential brown dwarf around the F star is, in reality, an M-dwarf star transiting the G star. 

The high brightness ratio of two stars posed significant challenges to revealing the true nature of the BD-14 3065 system. To do so, we adopted the approach detailed in \cite{Mandushev05} on the TRES spectra. Initially, we computed the bisector spans (BIS) of cross-correlation functions (CCFs) to examine the asymmetries of spectral lines. We then determined the periodogram of the BIS time series (see Section \ref{sec:Freq}), which revealed two peaks at 4.3 and 3.3 days. These peaks indicated that two CCFs are moving with respect to each other with the period of orbital motion, causing asymmetries in the blended CCF. However, here, we observed a key difference between the GSC01944-02289 and BD-14 3065 systems. While BIS showed a semiamplitude of more than one $\mathrm{km\,s^{-1}}$ in the case of GSC01944-02289, BD-14 3065 BIS only showed a semiamplitude of less than 300 $\mathrm{m\,s^{-1}}$, and the peak was barely significant, albeit visible.

\begin{figure*}
\centering
\includegraphics[width=1.0\textwidth,height=0.8\textwidth]{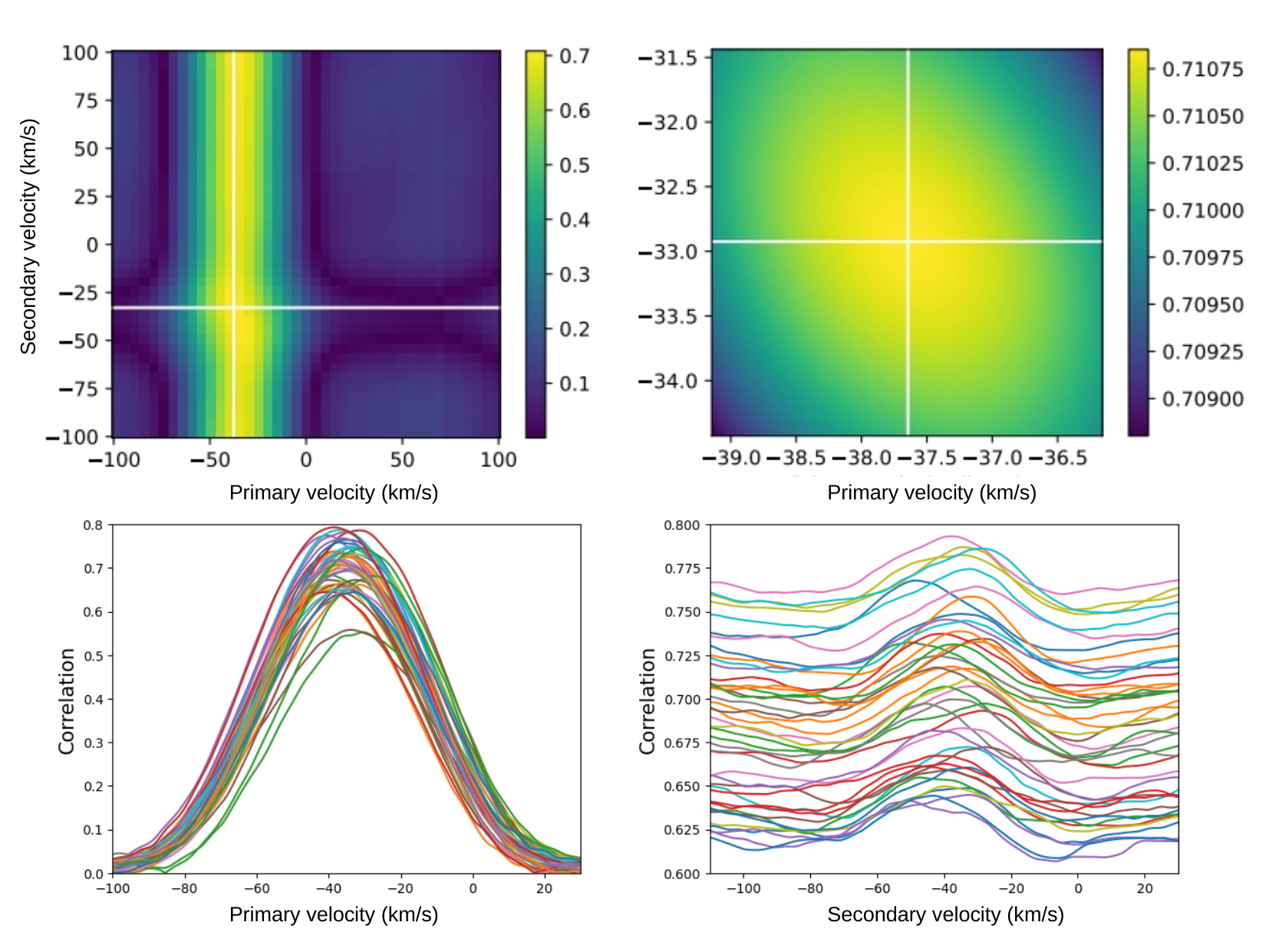}
\caption{Upper panels: Two-dimensional cross-correlation surface computed with TODCOR as a function of the F and G star velocities for one spectrum. The white lines represent the adopted velocities for both stars. Bottom panels: Cross-sections of the two-dimensional correlation functions at a fixed velocity for the G star and F star, respectively.}
\label{fig:todcor}
\end{figure*}

Subsequently, we performed the two-dimensional cross-correlation algorithm TODCOR \citep{Zucker94}. TODCOR cross-correlates the observed spectra against a composite template obtained by adding together two distinct templates chosen to match each star. We used stellar parameters from the two-star SED modeling and assumed that the stars in the potential eclipsing binary would likely have their rotation synchronized with the orbital motion due to tidal forces (translating to the vsini of 11 $\mathrm{km\,s^{-1}}$). We were able to measure the RVs of both stars. An illustration of this detection is given in Fig. \ref{fig:todcor}, which shows the two-dimensional correlation function as well as cross-sections of the two-dimensional correlation function at a fixed velocity for the G star and F star, respectively. 

Notably, the detection of the transiting object around the F star (see Figure \ref{fig:per}) confirms its substellar nature.

To further validate the results from the TODCOR analysis, we conducted additional tests. Firstly, we allowed the light fraction factor to be a free parameter, and the results indicated that the best value was close to the expected value of 0.1 from speckle imaging. Secondly, we analyzed data of BIS from the cross-section of 2D CCF of the F-type star and computed its periodogram. This revealed that the magnitude of the asymmetries in the BIS data decreased. It also eliminated the peak at the period of orbital motion, suggesting that we had successfully extracted the contamination of the G-type star. Importantly, the Pucheros+ data are less affected by the wide stellar companion than TRES data. Therefore, the presence of the signal of the transiting planet in this dataset indicates that it originates in the primary star. Finally, we compared the photometry of GSC01944-02289 and BD-14 3065. In the TESS LCs of GSC01944-02289, clear variations with the period of half the orbital period of the eclipsing star are visible, with maxima in phases 0.25 and 0.75. The stellar deformation of close binary stars causes these variations. We would, therefore, expect to observe similar variations for BD-14 3065. However, the LCs instead show variations with a period similar to the transiting companion's period and a clear evolution of this variability through TESS sectors, which we interpret as the rotation period of the F-type star (see Fig. \ref{fig:tess_lc}). It is noteworthy that most of the flux in the TESS photometry is coming from the F star, and if these photometric variations originated from the G star, they would be suppressed by the flux from the F star. It would be surprising for an old G-type star to exhibit such large fluctuations. If the companion orbited the G star and no tidal interactions were in play, the close rotation period of the F star would be an unlikely coincidence. 

In conclusion, using the TODCOR tool, we were able to distinguish F and G star components in the cross-correlation function. Our additional validations revealed that the analysis was successful and the transiting object is around the F star. Moreover, the signal of the transiting planet is also visible in less contaminated Pucheros+ data and our discussion suggests that the TESS photometry aligns with this conclusion.

As an alternative approach, ground-based photometry may be utilized to differentiate which star the transiting body orbits. Our calculations indicate that if the body orbits the primary F star, transit depths in the Sloan $g'$ filter are expected to be only 0.1\,ppt deeper than in the Sloan $i'$ filter. However, such a small difference would not be distinguishable in the LCOGT photometry. In the event that the body transits one of a fainter star, transit depths in the Sloan $i'$ filter would be roughly 1\,ppt deeper than in the Sloan $g'$ filter. With applied up to two detrending vectors to the LCOGT photometry, we obtained a good solution for both scenarios, one in which transit depths are similar in both filters and another with transit depth shallower in the Sloan $g'$ filter. Due to the limited number of out-of-transit observations, we were unable to investigate the impact of systematics on the transit depth. Nevertheless, by computing the Bayesian model log evidence (ln\,Z), a widely used metric, we can distinguish the scenario with the same depth transits as significantly better. In Figure \ref{fig:phot_all} we plot the scenario with all transists having a similar depth.

\subsection{Frequency Analysis}\label{sec:Freq}

\begin{figure}
\centering
\includegraphics[width=0.46\textwidth, trim= {0.0cm 0.0cm 0.0cm 0.0cm}]{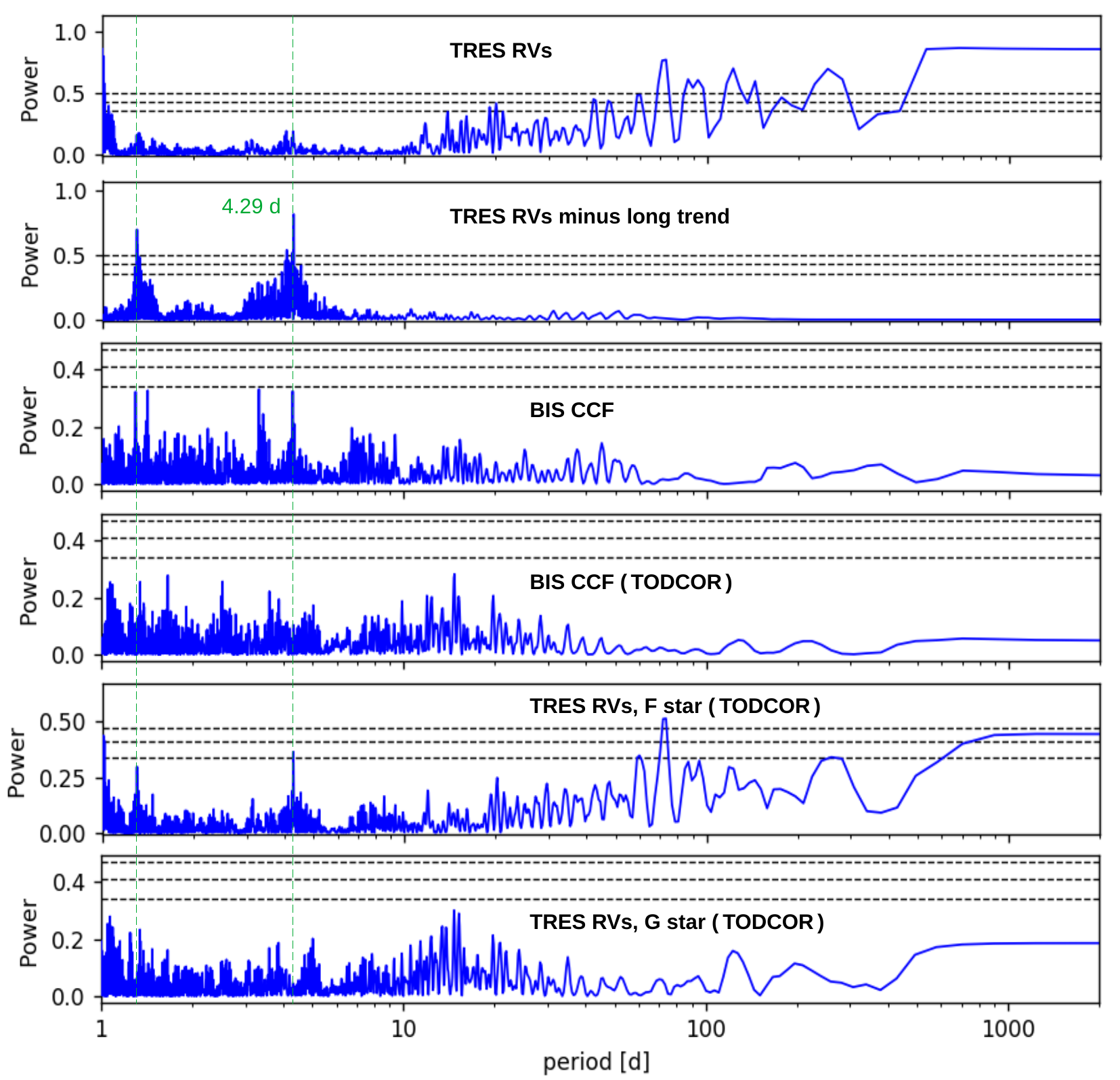}
\caption{Generalized Lomb-Scargle periodograms of TRES RVs of BD-14\,3065: (a) TRES RVs, (b) TRES RVs minus long trend, (c) TRES BIS CCF, (d) TRES BIS CCF from TODCOR, (e) TRES RVs of the F star from TODCOR, (f) TRES RVs of the G star from TODCOR. The vertical green lines highlight the orbital period of the transiting companion and its 1-day alias. Horizontal dashed lines show the theoretical FAP levels of 10\%, 1\%, and 0.1\% for each panel.} \label{fig:per}
\end{figure}

We performed a frequency analysis to detect all signals in RVs. Our observations had a baseline of around 500 days, providing a frequency resolution of approximately $0.002d^{-1}$. We observed a non-linear long trend in the dataset, with an unconstrained period as seen in the periodogram. This signal corresponds to the close stellar companion, which we accounted for in the three-component SED modeling. The period, eccentricity, and mass of the companion remain unconstrained, although its minimum mass is in the stellar regime (assuming the shortest possible period). After fitting the trend, we identified a significant period of 4.3 days and its one-day alias, consistent with the transits from TESS photometry (see Fig. \ref{fig:per}).


To test the effect of the unconstrained period/mass of the close stellar companion BD-14\,3065c on the parameters of the transiting planet BD-14\,3065b, we gradually increased the mass of the companion from a low-mass M dwarf star to a G-type star (the result from a blended three-component SED model). We initialized the MCMC run for each mass using {\tt exostriker} \citep{Trifonov19} to derive posterior distributions of the transiting planet parameters. We conclude that these parameters are well-constrained and consistent within their error bars, regardless of the mass of the close stellar companion.

\subsection{Joint RVs, transits and eclipses modelling}\label{sec:RVs}

The process of modeling transits, eclipses, and radial velocities was carried out simultaneously using the {\tt allesfitter} package. The fitting routine, as described in \cite{Gunther21}, is based on several packages, including {\tt ellc} for light-curve and RV models \citep{Maxted16}, {\tt aflare} for flare modeling \citep{Davenport14}, {\tt dynesty} for static and dynamic Nested Sampling \citep{Speagle20}, {\tt emcee} for MCMC sampling \citep{Foreman13}, and {\tt celerite} for GP models \citep{Foreman17}. The {\tt allesfitter} package offers a range of orbital and transit/eclipse models, accommodating multiple exoplanets, multi-star systems, transit-timing variations, phase curves, stellar variability, starspots, stellar flares, and various systematic noise models, including Gaussian processes.

\begin{figure*}
\centering
\includegraphics[width=1.0\textwidth]{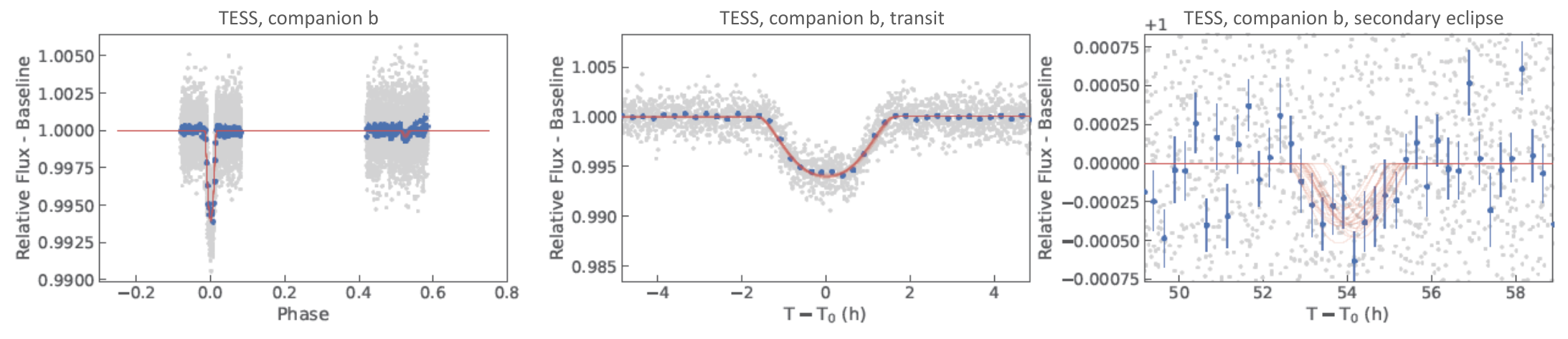}
\caption{The phased LC of the BD-14\,3065, fitted with the {\tt allesfitter} as described in Section \ref{sec:RVs}. The grey points represent TESS data, and the blue points are TESS binned data in the phase curve with a bin width of 15 minutes in the phase. The red line represents the best transit and eclipse model.} \label{fig:secondary}
\end{figure*}

To obtain the optimal conjunction time, period, inclination, eccentricity, and argument of periastron, which were later used as priors for joint modeling, we performed a fit of photometry alone, as the secondary eclipse is visible in TESS photometry (see Fig. \ref{fig:secondary}). The LCs were already corrected for variability using the procedure outlined in Section \ref{sec:TESS}, and we employed the quadratic limb darkening law with coefficients q1 and q2. To ensure accuracy and avoid degeneracy between these coefficients and transit depth, we adopted Gaussian priors within $0.1\,\sigma$ around the values from \cite{Claret17}, which presents limb-darkening coefficients specifically computed for the TESS photometry for a wide range of effective temperatures, gravities, metallicities, and microturbulent velocities. Although modeling the pulsations simultaneously with the transits may be preferred, it will not impact the results since pulsations are much slower than the transit duration. We conducted two separate fits based on the stellar parameters and dilution factors obtained from two-star and three-star SED modeling.  

We conducted a joint fit of our data, incorporating TRES and Pucheros+ RVs processed in accordance with the procedures outlined in Section \ref{sec:observations}. In the case of the TRES data, we did not use RVs obtained from the TODCOR analysis, as we discovered that the signal is associated with the F star, rendering any significant difference unlikely. To test this assertion, we modeled the orbit using RVs obtained from the TODCOR analysis in Section \ref{sec:observations}, and we found that all parameters were consistent within $1\sigma$. Additionally, the RVs used in Section \ref{sec:observations} were obtained via a more sophisticated method, utilizing the multi-order approach.
Table \ref{tab:sec_fit} shows the priors and posteriors of fitted parameters, while Table \ref{tab:sec_der} presents the transit parameters and physical parameters derived from Tables \ref{tab:compare} and \ref{tab:sec_fit}, respectively. The results are plotted in Fig. \ref{fig:secondary}.

The stellar radius is an important parameter that leads to the companion radius computed through the transit depth. We need to double-check that the radius of BD-14\,3065b is correct, as the discussion of this work is built upon its value. As a first check, we modeled the 2-star SED, assuming that the star found in the high-resolution imaging is an unassociated companion with its own age and distance. However, such a model led to similar parameters of the primary star ($M_\star=1.38\pm0.03$\,M$_{\odot}$, a radius of $R_\star=2.40\pm0.08$\,R$_{\odot}$). As a second check, to ensure the reliability of reported stellar spectral parameters, which are crucial in our SED modeling, we conducted a search for values in the relevant literature. The Sixth Data Release of the Radial Velocity Experiment (RAVE) reported stellar parameters of $T_{\rm eff}=6809\pm140$\,K and $\log{g}=3.78\pm0.19$. Gaia reported stellar parameters of $T_{\rm eff}=6791\pm32$\,K and $\log{g}=3.90\pm0.01$ using the photometry\footnote{\url{https://gea.esac.esa.int/archive/documentation/GDR3/Data_analysis/chap_cu8par/sec_cu8par_apsis/ssec_cu8par_apsis_gspphot.html}}. These values are consistent with our own findings. Additionally, it is important to note that unresolved binarity is not expected to impact these values significantly \citep{ElBadry18}. We are, therefore, confident that the reported stellar parameters are reliable. Finally, the impact of an unresolved binary star on the radius of a transiting planet has been the subject of much discussion in the literature. The ratio of the planet's true radius $R_{P,\,true}$, to the observed radius $R_{P,\,observed}$ when assuming a single star with no companions, can be described as:

\begin{equation}\label{radius_correction}
      X_R \equiv {\frac{R_{P,\,true}}{R_{P,\,observed}}} \equiv
         \left(\frac{R_{\star,\,t}}{R_{\star,\,1}}\right)\,\sqrt{\frac{F_{total}}{F_{t}}}
   \end{equation}

\noindent
where $R_{\star,\,1}$ represents the radius of the primary star (assumed to be a single star), and $F_{t}$ and $R_{\star,\,t}$ represent the brightness and radius, respectively, of the star that is being transited by the planet. When the true stellar radius is determined, unaffected by stellar companions, for example, through asteroseismic analysis or spectroscopic analysis together with stellar isochrones, we have $R_{\star,\,1} \equiv R_{\star,\,t}$ for a planet transiting the primary star \citep{Hirsch17}. In our study, we derived stellar parameters from SED modeling, and ${R_{\star,\,1}}>{R_{\star,\,t}}$. However, assuming that our effective temperature is not largely underestimated, $\frac{R_{\star,\,1}}{R_{\star,\,t}}\sim\sqrt{\frac{F_{total}}{F_{t}}}$, and we do not anticipate an overestimation of the radius of BD-14 3065b.

\begin{figure*}
\centering
\includegraphics[width=1.0\textwidth]{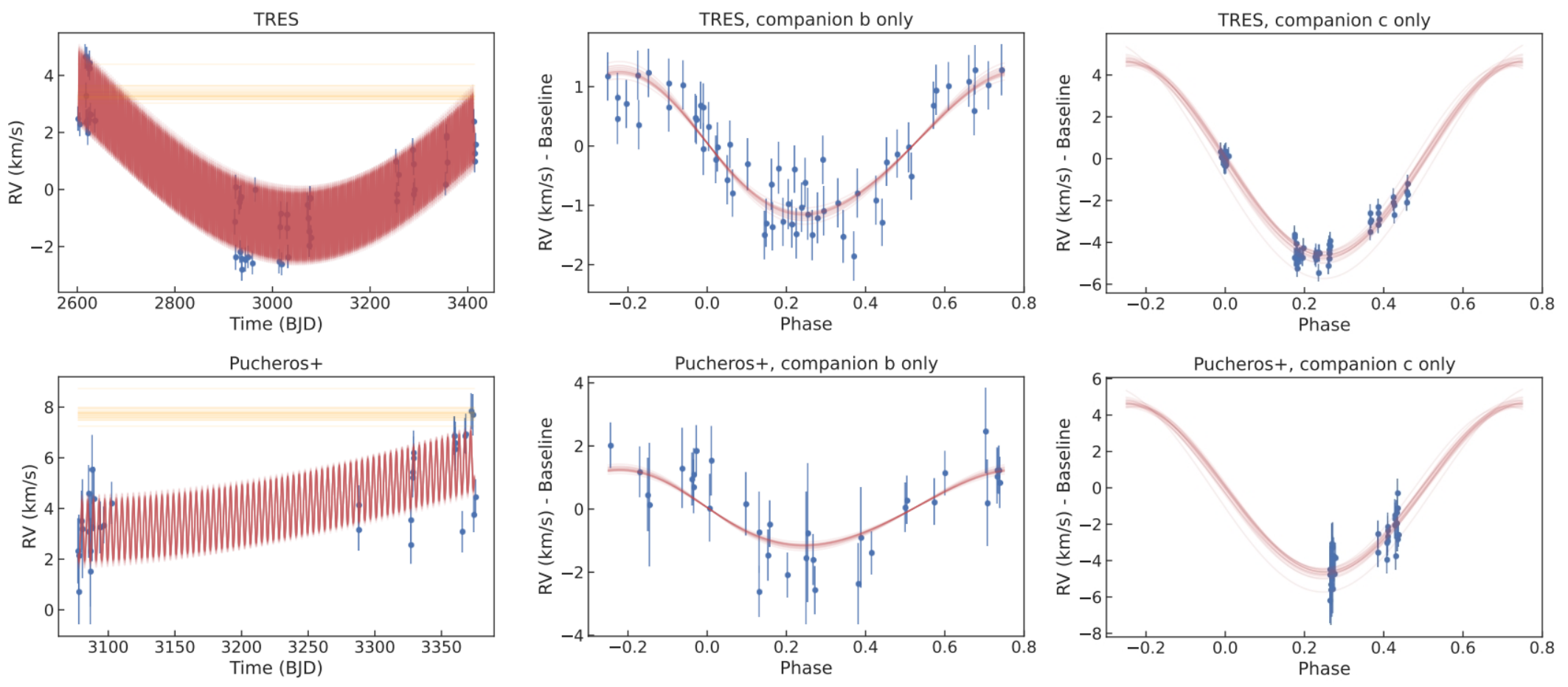}
\caption{The radial velocity curve of BD-14\,3065, fitted with the {\tt allesfitter} as described in Section \ref{sec:RVs}. The blue points represent TRES and Pucheros+ data, and the red lines represent the best models that result from the same global fit of both datasets together. The decision to plot both datasets separately was based on the difference in their average uncertainties. For illustration purposes, we have plotted the shortest feasible circular orbit for BD-14\,3065c. Hence, we have plotted a two-star model that assumes that the stellar companion BD-14\,3065c does not cause significant dilution. We have subtracted a value of 2,457,000 from BJD time.} \label{fig:RVs}
\end{figure*}

\begin{table}
 \centering
 \caption{Fitted parameters from the {\tt allesfitter} analysis.} 
 \label{tab:sec_fit}
 \begin{tabular}{lccr}
    \hline
    \hline
    Parameter & Value & Unit & fit/fixed \\
    \hline
    \multicolumn{4}{c}{\textit{Fitted parameters: 2-star model}} \\ 
    \hline 
    $R_b / R_\star$ & $0.0843_{-0.0022}^{+0.0023}$ & & fit \\ 
    $(R_\star + R_b) / a_b$ & $0.1804_{-0.0028}^{+0.0032}$ & & fit \\ 
    $\cos{i_b}$ & $0.1602_{-0.0033}^{+0.0036}$ & & fit \\ 
    $T_{0;b}$ & $2313.48578\pm0.00076$ & $\mathrm{BJD}$& fit \\ 
    $P_b$ & $4.2889731\pm0.0000052$ & $\mathrm{d}$& fit \\ 
    $K_b$ & $1.210\pm0.071$ & $\mathrm{km\,s^{-1}}$& fit \\ 
    $\sqrt{e_b} \cos{\omega_b}$ & $0.152\pm0.013$ & & fit \\ 
    $\sqrt{e_b} \sin{\omega_b}$ & $0.208_{-0.033}^{+0.024}$ & & fit \\ 
    $D_\mathrm{0; TESS}$ & $0.106\pm0.010$ & & fixed \\ 
    $q_{1; \mathrm{TESS}}$ & $0.111_{-0.110}^{+0.073}$ & & fit \\ 
    $q_{2; \mathrm{TESS}}$ & $0.209\pm0.093$ & & fit \\ 
    $J_{b; \mathrm{TESS}}$ & $0.176_{-0.054}^{+0.070}$ & & fit \\ 
    ${\sigma_\mathrm{TESS}}$ & $1332\pm11$ & ${ \mathrm{ppm} }$& fit \\ 
    ${\sigma_\mathrm{jitter; TRES}}$ & $0.32\pm0.05$ & ${ \mathrm{km\,s^{-1}} }$& fit \\  
    \hline
    \multicolumn{4}{c}{\textit{Fitted parameters: 3-star model}} \\
    \hline 
    $R_b / R_\star$ & $0.0901\pm0.0022$ & & fit \\ 
    $(R_\star + R_b) / a_b$ & $0.1711_{-0.0029}^{+0.0031}$ & & fit \\ 
    $\cos{i_b}$ & $0.1495\pm0.0038$ & & fit \\ 
    $T_{0;b}$ & $2313.48571\pm0.00078$ & $\mathrm{BJD}$& fit \\ 
    $P_b$ & $4.2889734\pm0.0000051$ & $\mathrm{d}$& fit \\ 
    $K_b$ & $1.200\pm0.071$ & $\mathrm{km\,s^{-1}}$& fit \\ 
    $\sqrt{e_b} \cos{\omega_b}$ & $0.148_{-0.013}^{+0.014}$ & & fit \\ 
    $\sqrt{e_b} \sin{\omega_b}$ & $0.220_{-0.036}^{+0.025}$ & & fit \\ 
    $D_\mathrm{0; TESS}$ & $0.226\pm0.10$ & & fixed \\ 
    $q_{1; \mathrm{TESS}}$ & $0.143_{-0.088}^{+0.098}$ & & fit \\ 
    $q_{2; \mathrm{TESS}}$ & $0.210\pm0.099$ & & fit \\ 
    $J_{b; \mathrm{TESS}}$ & $0.160_{-0.049}^{+0.067}$ & & fit \\     
    ${\sigma_\mathrm{TESS}}$ & $1330\pm11$ & ${ \mathrm{ppm} }$& fit \\
    ${\sigma_\mathrm{jitter; TRES}}$ & $0.32\pm0.05$ & ${ \mathrm{km\,s^{-1}} }$& fit \\ 
    \hline
    \hline
    \end{tabular}
    Comments: a -- BJD - 2,457,000, b -- dilution factor, c -- surface brightness ratio between the companion and star
\end{table} 

\begin{table}
 \centering
 \caption{Derived parameters from the {\tt allesfitter} analysis.} \label{tab:sec_der}
 \begin{adjustbox}{width=0.48\textwidth}
 \begin{tabular}{lc}
    \hline
    \hline
    Parameter & Value \\
    \hline
    \multicolumn{2}{c}{\textit{Derived parameters: 2-star model}} \\ 
    \hline
    Host radius over semi-major axis b; $R_\star/a_\mathrm{b}$ & $0.1664_{-0.0032}^{+0.0037}$\\ 
    Semi-major axis b over host radius; $a_\mathrm{b}/R_\star$ & $6.01_{-0.13}^{+0.12}$\\ 
    Companion radius b over semi-major axis b; $R_\mathrm{b}/a_\mathrm{b}$ & $0.01403_{-0.00047}^{+0.00060}$\\ 
    Companion radius b; $R_\mathrm{b}$ ($\mathrm{R_{\oplus}}$) & $21.59\pm1.05$\\ 
    Companion radius b; $R_\mathrm{b}$ ($\mathrm{R_{jup}}$) & $1.926\pm0.094$\\ 
    Semi-major axis b; $a_\mathrm{b}$ ($\mathrm{R_{\odot}}$) & $14.11\pm0.57$\\ 
    Semi-major axis b; $a_\mathrm{b}$ (AU) & $0.0656\pm0.0026$\\ 
    Inclination b; $i_\mathrm{b}$ (deg) & $80.78_{-0.30}^{+0.24}$\\ 
    Eccentricity b; $e_\mathrm{b}$ & $0.0660_{-0.011}^{+0.0098}$\\ 
    Argument of periastron b; $w_\mathrm{b}$ (deg) & $54.0_{-9.1}^{+6.0}$\\ 
    Mass ratio b; $q_\mathrm{b}$ & $0.00838_{-0.00052}^{+0.00055}$\\ 
    Companion mass b; $M_\mathrm{b}$ ($\mathrm{M_{\oplus}}$) & $3932_{-280}^{+290}$\\ 
    Companion mass b; $M_\mathrm{b}$ ($\mathrm{M_{jup}}$) & $12.37_{-0.87}^{+0.92}$\\ 
    Companion mass b; $M_\mathrm{b}$ ($\mathrm{M_{\odot}}$) & $0.01181_{-0.00083}^{+0.00088}$\\ 
    Impact parameter b; $b_\mathrm{tra;b}$ & $0.9099_{-0.0079}^{+0.0092}$\\ 
    Total transit duration b; $T_\mathrm{tot;b}$ (h) & $3.086\pm0.041$\\ 
    Full-transit duration b; $T_\mathrm{full;b}$ (h) & $0.61_{-0.22}^{+0.19}$\\ 
    Epoch occultation b; $T_\mathrm{0;occ;b}$ & $2315.7361_{-0.0074}^{+0.0084}$\\ 
    Impact parameter occultation b; $b_\mathrm{occ;b}$ & $1.013_{-0.024}^{+0.020}$\\ 
    Host density from orbit b; $\rho_\mathrm{\star;b}$ (cgs) & $0.223_{-0.014}^{+0.013}$\\ 
    Companion density b; $\rho_\mathrm{b}$ (cgs) & $2.15_{-0.48}^{+0.53}$\\ 
    Companion surface gravity b; $g_\mathrm{b}$ (cgs) & $10536\pm950$\\ 
    Equilibrium temperature b; $T_\mathrm{eq;b}$ (K) & $2001\pm31$\\ 
    Transit depth (undil.) b; $\delta_\mathrm{tr; undil; b; TESS}$ (ppt) & $6.49_{-0.20}^{+0.23}$\\ 
    Transit depth (dil.) b; $\delta_\mathrm{tr; dil; b; TESS}$ (ppt) & $5.80_{-0.11}^{+0.14}$\\ 
    Occultation depth (undil.) b; $\delta_\mathrm{occ; undil; b; TESS}$ (ppt) & $0.482_{-0.11}^{+0.091}$\\ 
    Occultation depth (dil.) b; $\delta_\mathrm{occ; dil; b; TESS}$ (ppt) & $0.431_{-0.096}^{+0.081}$\\ 
    \hline
    \multicolumn{2}{c}{\textit{Derived parameters: 3-star model}} \\ 
    \hline 
    Host radius over semi-major axis b; $R_\star/a_\mathrm{b}$ & $0.1570_{-0.0027}^{+0.0028}$\\ 
    Semi-major axis b over host radius; $a_\mathrm{b}/R_\star$ & $6.37\pm0.11$\\ 
    Companion radius b over semi-major axis b; $R_\mathrm{b}/a_\mathrm{b}$ & $0.01414_{-0.00045}^{+0.00047}$\\ 
    Companion radius b; $R_\mathrm{b}$ ($\mathrm{R_{\oplus}}$) & $21.41\pm1.05$\\ 
    Companion radius b; $R_\mathrm{b}$ ($\mathrm{R_{jup}}$) & $1.910\pm0.094$\\ 
    Semi-major axis b; $a_\mathrm{b}$ ($\mathrm{R_{\odot}}$) & $13.88\pm0.57$\\ 
    Semi-major axis b; $a_\mathrm{b}$ (AU) & $0.0646\pm0.0026$\\ 
    Inclination b; $i_\mathrm{b}$ (deg) & $81.40\pm0.22$\\ 
    Eccentricity b; $e_\mathrm{b}$ & $0.0702_{-0.011}^{+0.0096}$\\ 
    Argument of periastron b; $w_\mathrm{b}$ (deg) & $56.2_{-7.2}^{+4.7}$\\ 
    Mass ratio b; $q_\mathrm{b}$ & $0.00826\pm0.00056$\\ 
    Companion mass b; $M_\mathrm{b}$ ($\mathrm{M_{\oplus}}$) & $3850_{-290}^{+300}$\\ 
    Companion mass b; $M_\mathrm{b}$ ($\mathrm{M_{jup}}$) & $12.11_{-0.91}^{+0.95}$\\ 
    Companion mass b; $M_\mathrm{b}$ ($\mathrm{M_{\odot}}$) & $0.01156_{-0.00087}^{+0.00091}$\\ 
    Impact parameter b; $b_\mathrm{tra;b}$ & $0.8959\pm0.0067$\\ 
    Total transit duration b; $T_\mathrm{tot;b}$ (h) & $3.052\pm0.042$\\ 
    Full-transit duration b; $T_\mathrm{full;b}$ (h) & $0.78_{-0.17}^{+0.14}$\\ 
    Epoch occultation b; $T_\mathrm{0;occ;b}$ & $2315.7365\pm0.0070$\\ 
    Impact parameter occultation b; $b_\mathrm{occ;b}$ & $1.007_{-0.024}^{+0.019}$\\ 
    Host density from orbit b; $\rho_\mathrm{\star;b}$ (cgs) & $0.266\pm0.014$\\ 
    Companion density b; $\rho_\mathrm{b}$ (cgs) & $2.16_{-0.31}^{+0.37}$\\ 
    Companion surface gravity b; $g_\mathrm{b}$ (cgs) & $10270\pm850$\\ 
    Equilibrium temperature b; $T_\mathrm{eq;b}$ (K) & $1957\pm28$\\ 
    Transit depth (undil.) b; $\delta_\mathrm{tr; undil; b; TESS}$ (ppt) & $7.42_{-0.12}^{+0.13}$\\ 
    Transit depth (dil.) b; $\delta_\mathrm{tr; dil; b; TESS}$ (ppt) & $5.741_{-0.092}^{+0.10}$\\ 
    Occultation depth (undil.) b; $\delta_\mathrm{occ; undil; b; TESS}$ (ppt) & $0.57\pm0.12$\\ 
    Occultation depth (dil.) b; $\delta_\mathrm{occ; dil; b; TESS}$ (ppt) & $0.442\pm0.089$\\ 
    \hline
    \hline
 \end{tabular}
 \end{adjustbox}
Comments: a -- computed assuming zero albedo
\end{table}

%
%

\section{Discussion} \label{sec:discussion}
\subsection{Mass-radius diagram}\label{sec:mass-radius}

\begin{figure*}
\centering
\includegraphics[width=1.0\textwidth]{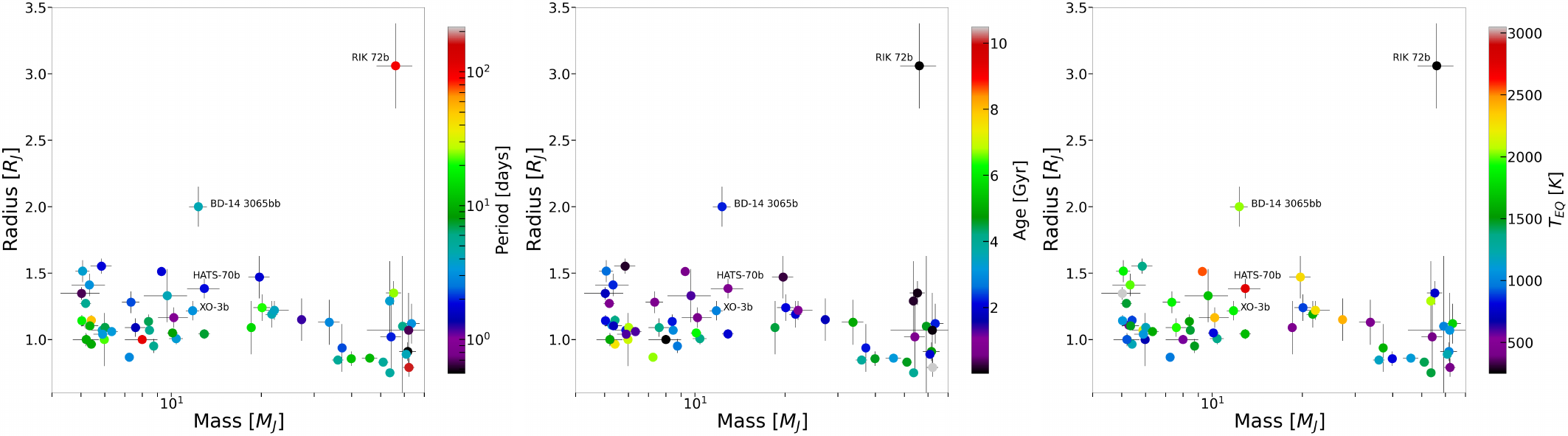}
\caption{The population of known companions within the mass interval $5-70\,\mj$. The position of BD-14\,3065b is highlighted together with the three objects mentioned in the discussion. The left subplot is colored with respect to the age of the systems, the middle one with respect to the orbital period of the planets, and the right one with respect to the equilibrium temperature of the planets.} \label{fig:M-R}
\end{figure*}

Combining data from the TESS mission and ground-based spectroscopy confirmed the substellar nature of a $P=4.29$\,days candidate around the F-type subgiant BD-14\,3065. We found that the physical parameters of BD-14\,3065b ($M_p=12.37\pm0.92\,\mj$, $R_p=1.926\pm0.094\,\rj$) are consistent with an object at the boundary between giant planets and brown dwarfs based on the deuterium burning criterium. A comparison of the mass and radius of BD-14\,3065b with the population of well-characterized brown dwarfs and giant planets with a mass greater than $5\,\mj$ is presented in Fig. \ref{fig:M-R}. Three subplots are created, with colors representing different parameters: age, orbital period, and equilibrium temperature. The position of BD-14\,3065b on the mass-radius diagram is exceptional. It is one of two objects with a significantly larger radius compared to the rest of the population. The second object is the brown dwarf RIK 72b \citep{David19} in the Upper Scorpius OB association, with an age of approximately 5\,Myrs. RIK 72b transits the pre-main sequence M-dwarf star with a period of $\sim98$\,days, and thus, most of its internal thermal energy originates from its contraction. The authors demonstrate that current evolutionary models that predict the size and luminosity of brown dwarfs as a function of age are consistent with the derived parameters. Conversely, BD-14\,3065 is an old star without lithium in the spectrum. Old age of about 2.3\,Gyr was derived using stellar isochrones for the subgiant primary star. Explaining the observed radius for such an old age is challenging using substellar evolutionary models, even when considering stellar irradiation, which slows down contraction.

The research conducted by \cite{Baraffe03} delved into the influence of stellar irradiation on the radius of giant planets with a maximum mass of $10\,\mj$. It is noteworthy that BD-14\,3065b is slightly more massive. However, the effect of irradiation decreases as the mass of the planet increases, given a certain incident flux. According to their models, for an object with a mass of about $10\,\mj$ and an age similar to that of BD-14\,3065, irradiation would inflate its radius by approximately $0.1\,\rj$. It is important to note that they hypothesized a lower stellar temperature of 6000\,K but also assumed a smaller irradiated planet separation of 0.046\,au. Even when considering the effects of stellar irradiation, the theoretical models can only match the radius of BD-14\,3065b for a few million years. To demonstrate this conclusion, we utilized the thermal evolution model outlined by \cite{Fortney07}, which also considers the impact of stellar irradiation. We assumed a core mass of $10\,M_{\oplus}$ and adjusted the model to BD-14\,3065’s luminosity and BD-14\,3065b’s orbital distance. We found only a small inflation of the radius; hence, the irradiation cannot replicate the observed radius of the companion.

In Fig. \ref{fig:M-R}, in the plot where colors represent object equilibrium temperature for a zero albedo, we can see that objects with larger equilibrium temperatures typically tend to have a higher radius. This suggests that the radius in many systems is affected by stellar irradiation. Equilibrium temperature refers to the theoretical temperature an object reaches when it is in thermal equilibrium, assuming it is only being heated by its parent star and following black body re-radiation. 

We can compare BD-14\,3065b with two systems hosting the same mass companions: HATS-70 \citep{Zhou19} and XO-3 \citep{JohnsKrull08}. HATS-70b is an object with a mass of $12.9\,\mj$ that transits a metallic-line A-type star with a period of 1.9 days. Due to its proximity to the star, it receives a higher amount of stellar radiation than BD-14\,3065b, resulting in an equilibrium temperature of 2750\,K. While the authors suggest that the stellar irradiation may account for the observed inflated radius of $1.38\,\rj$, they note that other well-documented effects may also play a role, such as enhanced atmospheric metallicity and opacity \citep[e.g.,][]{Burrows07,Burrows11} or ohmic dissipation \citep[e.g.,][]{Batygin10}. XO-3b is an object with a mass of $11.8\,\mj$ that transits an F-type star with a period of 3.2\,days. \cite{Dang22} employed Bayesian analysis to determine the amount of additional heating required to explain its radius of $1.3\,\rj$ as a fraction of the incident flux. The results indicate that approximately 20\% of additional heating from an external source is necessary, and the authors discussed potential sources such as tidal heating or deuterium burning. Additionally, \cite{Yang22} identified evidence of XO-3b's tidal evolution through transit timing variations, suggesting that tidal heating could be a source of additional heating. 

The inflated radii of many hot Jupiters have been attributed to several mechanisms, the intensity of which is often scaled to explain the observed radii. However, even among the population of inflated gas giants and brown dwarfs, BD-14\,3065b stands out with its unusually large radius for such a massive and old object. The potential reasons behind the unusually strong intensity of these mechanisms for BD-14\,3065b, or whether there is an unusual combination of several mechanisms at play, are beyond the scope of this work. Instead, we focus on the two most obvious differences between BD-14\,3065b and other systems in the literature. Firstly, BD-14\,3065b is situated at the boundary of deuterium burning, and secondly, its primary star is an evolved subgiant.

\subsection{Deuterium burning}\label{sec:burn}

It is generally accepted that objects with masses below $13\,\mj$ are incapable of burning deuterium or can only do so to a small extent and are therefore classified as planets. \cite{Moliere12} developed a model to determine the initial deuterium burned based on object mass, helium fraction, metallicity, gas accretion rate, and deuterium abundance. According to their modeling, BD-14\,3065b may still retain a significant portion of its deuterium despite the system's age. Hence, deuterium represents an energetic reservoir inside the object. Additionally, their model suggests that low-mass objects, such as BD-14\,3065b, continue to burn a small fraction of deuterium over several Gyr, following a strong burning phase that lasts approximately 100\,Myr. The rate of deuterium reaction is highly dependent on the interior temperature, as expressed by the equation $\epsilon\,\propto\,[D/H]\,{\rho}\,T^{11.8}$. Therefore, increasing the object's interior temperature through tidal interactions with an evolving post-main-sequence star could lead to an increase in the rate of deuterium burning. 

\cite{Vissapragada22} observed evidence of tidally-driven inspiral of the Kepler-1658b planet around an evolved subgiant star, similar to BD-14\,3065, with an orbital period of 3.85 days, which is not far from BD-14\,3065b's period. Furthermore, using a secondary eclipse from Kepler data, they suggest that the planet is tidally superheated, with approximately ten percent of the energy from the shrinking orbit being dissipated in the planet. WASP-12b is another planet that has been observed to experience orbital decay \citep{Bailey19,Yee20}. The planet is orbiting an F-type star, which could be a subgiant. The planet has a highly inflated radius, and its brightness temperature has been determined to be $3640 \pm 230$\,K through secondary eclipse observations \citep{Crossfield12}. This temperature is notably higher than the planet's equilibrium temperature of $2990 \pm 110$\,K \citep{Crossfield12}. This suggests that the planet may be experiencing tidal superheating, similar to Kepler-1658b.

\begin{figure}
\centering
\includegraphics[width=0.46\textwidth, trim= {0.0cm 0.0cm 0.0cm 0.0cm}]{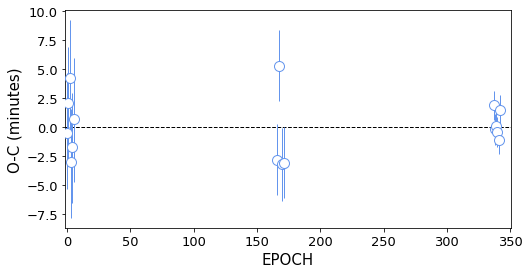}
\caption{Transit timing variations of BD-14\,3065b.} \label{fig:ttvs}
\end{figure}

Tidal evolution should eventually lead to planet engulfment if the orbital angular momentum is too low to allow the star to achieve spin-orbit synchronization \citep{Hut80}. However, using TESS photometry, we have determined that BD-14\,3065b is synchronized with the rotation of its host star. Hence, the current tidally-driven inspiral of the planet is not expected. To confirm this, we studied the transit timings of BD-14\,3065b with {\tt juliet} \citep{Espinoza19} and found no evidence of significant evolution. We note that the time sampling of 1800\,s in TESS sector 9 and 600\,s in sector 35 results in large uncertainties in the derived transit middle times. TTVs are shown in Fig. \ref{fig:ttvs}. It is also important to note that the radius of Kepler-1658b is not comparable to that of BD-14\,3065b.

In order to investigate the possibility of overheating in BD-14\,3065b, we derived its thermal emission through secondary eclipse depth analysis. The depth is proportional to the square of the radius ratio and the temperature ratio of the host and transiting companion. In Section \ref{sec:RVs}, we analyzed the TESS LCs and found a secondary eclipse depth of $\delta_S\,=\,0.48 \pm 0.11$ parts per thousand (ppt). The secondary eclipse fit can be observed in Fig. \ref{fig:secondary}. It is important to note that the TESS bandpass window\footnote{\url{https://archive.stsci.edu/files/live/sites/mast/files/home/missions-and-data/active-missions/tess/_documents/TESS_Instrument_Handbook_v0.1.pdf}} is centered at roughly 800\,nm and spans $600-1000$\,nm, which means that the relative brightness of the planet to the star is only measured at these wavelengths. We used this relative brightness to determine the effective temperature of BD-14\,3065b, employing the equation used for the secondary eclipse depth from \cite{Charbonneau05}:

\begin{equation}
    \delta_{S}\approx\bigg(\frac{R_b}{ R_\star}\bigg)^2\frac{\int F_b(\lambda)\,S(\lambda)\,\lambda\,d\lambda}{\int F_\star(\lambda)\,S(\lambda)\,\lambda\,d\lambda}
\end{equation}

\noindent
where $S(\lambda)$ is the TESS spectral response function and $F(\lambda)$ is the flux of companion or star. In order to derive BD-14\,3065b's effective temperature, we simplify the equation by assuming blackbody profiles to describe fluxes in the TESS bandpass window and zero geometric albedo. Our findings indicate that BD-14\,3065b has an effective blackbody temperature of $T_{eff}\,=\,3520\pm130\,K$. This temperature is notably higher than the equilibrium temperature that would result from star heating alone. Consequently, the secondary eclipse has revealed that the planet is experiencing overheating, which would increase the deuterium burning rate. The validity of this conclusion is contingent upon the assumption of BD-14\,3065b's blackbody spectrum. However, it remains uncertain whether the eclipse spectra of the hottest atmospheres align with blackbody curves. Studies of individual planets with HST and Spitzer have provided observational evidence in support of proximity to blackbody \citep[e.g.,][]{Kreidberg18,Mansfield18,Arcangeli19}. Nevertheless, recent research has highlighted a rise in brightness for wavelengths shorter than 1.3\,$\mu$m due to the combined opacities of H$^-$, TiO, VO, and FeH \citep[e.g.,][]{Changeat22,Coulombe23}. On the other hand, according to observations from \cite{Coulombe23}, TESS observation of planet thermal emission aligns well with the black body spectrum. Consequently, we are of the opinion that our assumption is reasonable.

Most models of deuterium burning consider the formation of objects through gravitational instability. Since such objects are both large and hot, deuterium burning can halt their contraction, which makes it challenging to evaluate the impact of deuterium burning on compact objects like BD-14\,3065b. \cite{Moliere12} modeled the effect of deuterium burning on objects formed through core accretion with a cold start, where gas is accreted with low entropy. Although this formation model is theoretical, it allows us to create an analogy with BD-14\,3065b. This model shows that deuterium burning in dense, compact objects would result in an increase in their radius, luminosity, and effective temperature. Definitive proof would require the modeling of the interior of BD-14\,3065b, taking into account all of the effects mentioned above. It is a difficult task that is beyond the scope of this work. 

The process of heating a planet's atmosphere can result in its expansion, but such a phenomenon does not fully account for the radii of many of the larger hot Jupiters. It is highly probable that planets with the largest observed radii have inflated interiors. The subsequent section will expound on a potential solution to the inflated radius of BD-14\,3065b by examining the effects of an overheated interior followed by increased deuterium burning.

\subsection{Overheating of BD-14\,3065b}\label{sec:overheating}

It is a well-established fact that the increase in entropy due to the interior heating of a planet leads to an increase in its radius \citep{Arras06,Marleau14}. In this regard, we aim to investigate the possibility of BD-14\,3065b being heated due to the interaction and dissipation between the star and the planet. Analysis of the TESS photometry data reveals that the stellar spin frequency is comparable with the planet's orbital mean motion, indicating that the synchronization timescale is comparable to the star's age. The synchronization timescale for the planet's spin angular frequency is smaller than the orbital synchronization timescale \citep{Bodenheimer01}. Therefore, the synchronization process of planet rotation cannot contribute significantly to the heating of the planetary interior.

Another potential heating source is the dissipation of a residual eccentricity, as indicated by a slightly eccentric orbit observed from a secondary eclipse. The rate of eccentric orbital energy dissipation inside the planet is proportional to the value of eccentricity, and this rate would decrease over time as the orbit becomes more circularized. The rate also depends on the circularization timescale or, more specifically, on the planet's and star's tidal quality factors, and while the circularization timescale is several orders smaller as stars become sub-giant, most of the eccentric orbital energy is dissipated inside the star rather than the planet. Therefore, while the planet may indeed be heated from a residual eccentricity, this source of heating weakens over time as the orbit becomes less eccentric. As this heating source was stronger in the past and affects many eccentric systems in literature, there is no obvious connection to BD-14\,3065's large radius. To calculate the rate of dissipation of the planet's eccentric orbital energy within the planet's interior, we used an equation from \cite{Bodenheimer01}. Assuming the planet's tidal quality factor $log\,Q' \approx 5.0$, we computed the value of $\dot{E}_{\rm D} \approx 3.6.10^{27}$\,erg\,s$^{-1}$.

Other possible sources could be dissipative processes powered by the stellar irradiation flux. These processes have received significant attention, yet the mechanism to transport a portion of the incident flux into the planet's interior remains an open question, and several processes have been proposed. One proposed mechanism is atmospheric circulation, which leads to the thermal dissipation of kinetic energy into the planet's interior \citep{Guillot02, Showman02}. Another proposed mechanism is ohmic dissipation, which is based on the idea that fast winds generated by the irradiation drive through the planet's magnetic fields to create currents that dissipate ohmically in the interior \citep[e.g.,][]{Batygin10, Huang12, Wu13, Ginzburg16}. The connection between dissipative processes and the evolutionary stage of the subgiant BD-14\,3065, which recently increased its radiation, makes these processes appear very realistic.

To compare the rate of energy transfer to the planet's interior to the one coming from an eccentric orbit and determine if the observed radius of BD-14\,3065b could be matched for a realistic fraction of stellar irradiation transported into the planet's interior, we used the hierarchical Bayesian code developed by \cite{Sarkis21}. The code utilizes the planetary evolution model {\tt completo21} \citep{Mordasini12} and fully non-gray atmospheric models from the {\tt petitCODE} \citep{Molliere15} to connect the interior structure models to the observed physical properties such as the planet's radius and mass, and the stellar luminosity and semimajor axis. The code assumes the planet is in a steady state and thus does not calculate the planetary thermal evolution. The planet models assume no central core, and the heavy elements (made up entirely of water) are equally distributed throughout the interior. A study by \cite{Sarkis21} found that the heating efficiency increases with increasing equilibrium temperature and reaches a maximum of 2.5\% at around 1860 K, beyond which the efficiency decreases. When assuming the highest heating efficiency of 2.5\%, we computed the value of $\dot{E}_{\rm T} \approx 2.3 \times 10^{28}$ erg s$^{-1}$, which is an order of magnitude higher than the energy dissipated from an eccentric orbit. However, this amount of energy can produce a radius only of about $1.2\,\rj$, assuming the stellar fraction of heavy element mass. Even if we assume a heating efficiency of 20\%, we can barely reach $1.6\,\rj$. There is no known process that can transfer such a high fraction of stellar irradiation into the interior of the planet. Therefore, we conclude that while several anticipated processes may increase the internal temperature of the planet, they would need to be unreasonably efficient to account for the observed radius. However, such processes would also significantly increase the rate of deuterium reactions. This appears to be the most plausible scenario. Alternatively, we can speculate that the planet may have experienced a recent collision that increased its internal temperature.

\begin{figure}
\centering
\includegraphics[width=0.46\textwidth, trim= {0.0cm 0.0cm 0.0cm 0.0cm}]{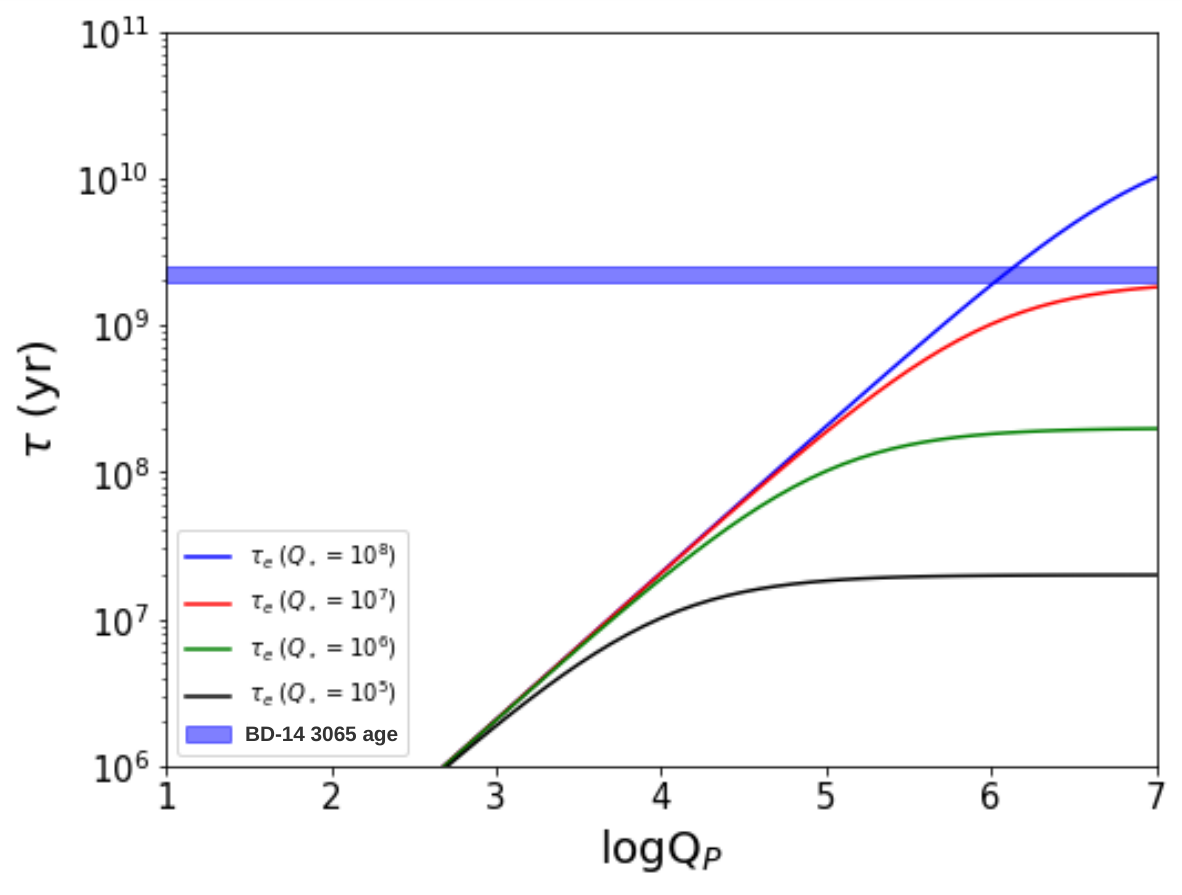}
\caption{Values of the tidal circularisation timescales as a function of planet tidal quality factor for BD-14\,3065b. Different colors represent the dependence on stellar tidal quality factors.} \label{fig:timescale}
\end{figure}

\begin{figure*}
\centering
\includegraphics[width=1.0\textwidth]{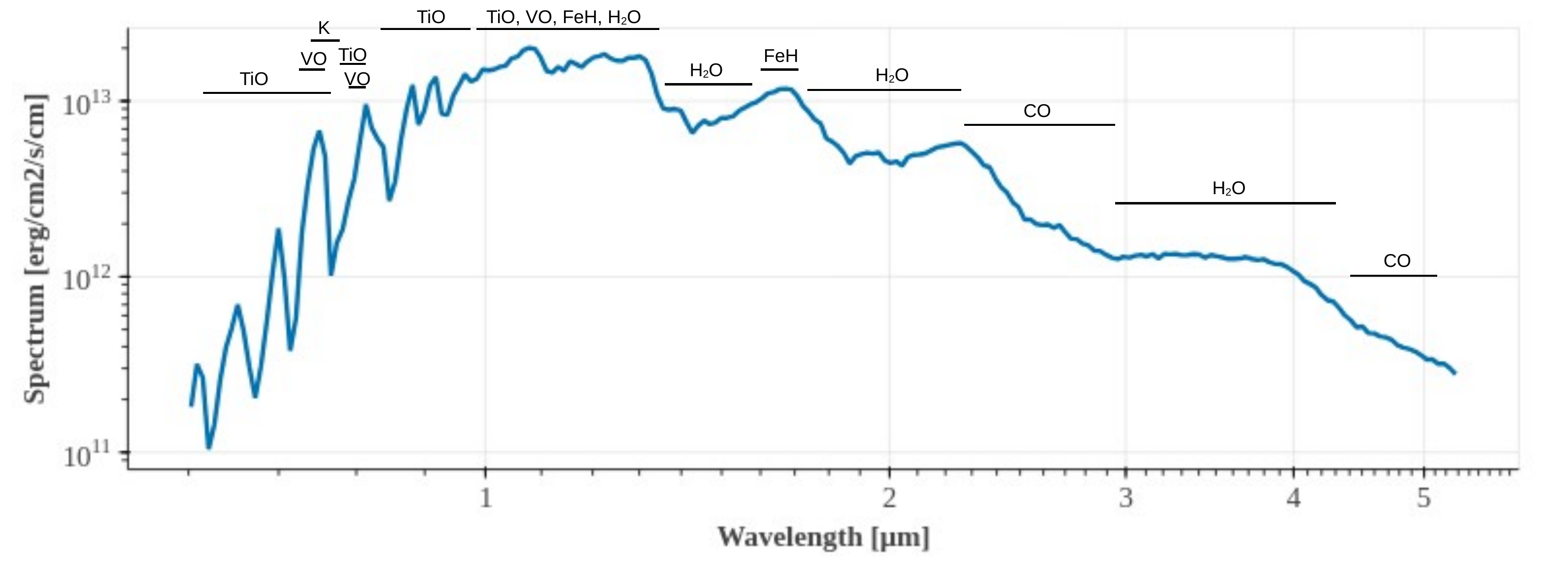}
\caption{Simulated emission spectrum of BD-14\,3065b as a function of wavelength. The model was generated by {\tt picaso}. The strongest atmospheric features are marked.} \label{fig:spec_b}
\end{figure*}

\subsection{Tidal interaction}\label{sec:tides}

Assuming that the observed eccentricity is residual from formation, an analysis of the circularization timescale can provide valuable insight into the star and planet tidal quality factors. Specifically, one may compare the circularization timescale with the age of the system. It is important to note that the circularization timescale represents the exponential decay of eccentricity rather than the time required to reach zero eccentricity.

As noted by \cite{Jackson2008}, the timescale required for orbital circularization of a close-in companion can be calculated using the following equation:

\begin{equation}
\label{eqn:tidecirc}
    \frac{1}{\tau_e}=\bigg[\frac{63}{4}\sqrt{GM_{\star}^3}\frac{R_{\mathrm{p}}^5}{Q'_{\mathrm{p}}M_{\mathrm{p}}}+\frac{171}{16}\sqrt{G/M_{\star}}\frac{R_{\star}^5M_{\mathrm{p}}}{Q'_{\star}}\bigg]a^{-\frac{13}{2}},
\end{equation}

\noindent
where $Q'_{\star}$ and $Q'_{\mathrm{p}}$ are the tidal quality factors of the star and planet, respectively. The tidal quality factor represents the ratio of the elastic energy stored to the energy dissipated per cycle of the tide, and larger values of the factor lead to longer circularization timescales. Equation \ref{eqn:tidecirc} further incorporates the Love number, the correction factor for the tidal-effective rigidity of the body and its radial density distribution, resulting in the expression $Q'=3Q/2k_{2}$ \citep{Goldreich66, Jackson2008}. For a homogeneous fluid body, $k_{2}=3/2$.

Measuring tidal quality factors is a challenging task. The number of brown dwarfs with published constraints on Q' is limited, and the available data indicate $log\,Q'_{\rm BD} \approx 4.15-4.5$ \citep{Heller10,Beatty18}. Numerous studies have been conducted to derive tidal quality factors for hot Jupiters, revealing typical values of $log\,Q'_{\rm J} \approx 5.0-6.5$ \citep[e.g.,][]{Jackson2008,Lainey2009,Quinn14}. There is a lack of consensus in the literature concerning the values of $Q'_\star$, which may range from $10^4$ to $10^8$. Furthermore, a single system's Q value can change over time due to the tidal forcing changes caused by both the system's orbital evolution and the evolution of the star \citep[e.g.,][]{Jackson2008,Penev12}.

In Figure \ref{fig:timescale}, we display the tidal circularisation timescale as a function of the tidal quality factors of the planet and star. It can be observed that within the typical range of $log\,Q'_{\rm J} \approx 5.0 - 6.5$, as reported in the literature, the circularisation timescale is only larger than system's age for the highest values of $Q'_\star$. This outcome aligns with the equilibrium tide values reported in the literature \citep{CollierCameron2018}, which should provide a reasonable estimate for a massive star like BD-14\,3065 \citep{Subjak20}. The tidal quality factor of subgiant stars, as measured by \cite{Vissapragada22}, is approximately $log\,Q'_{\star} \approx 4$, and for such a value, the circularisation timescale would decrease to a few million years.

\subsection{Follow-up prospects}\label{sec:follow-up}

BD-14\,3065b presents itself as a highly promising target for further atmospheric characterization through secondary eclipse observations. The planet's equilibrium temperature classifies it as an ultra-hot object that is comparable to WASP-18b, whose thermal emission spectrum was previously observed using the James Webb Space Telescope \citep{Coulombe23}. The most prominent features that can be observed in such an atmosphere include H$_2$O, CO, FeH, TiO, and VO absorption. A simple atmospheric model of the anticipated thermal emission spectrum in the NIRSpec and NIRISS wavelength ranges was created (see Fig. \ref{fig:spec_b}) using the {\tt picaso} code \citep{Batalha19} alongside the {\tt Sonora} brown dwarf evolutionary models \citep{Marley21}. In a similar fashion to WASP-18b, the James Webb Space Telescope would be able to determine the metallicity and carbon-to-oxygen (C/O) ratio of BD-14\,3065b.

C/O ratios have been measured for several transiting planets \citep[e.g.,][]{Madhusudhan11,Moses13,Line14,Arcangeli18,Ahrer23}. \cite{Beatty18} conducted a study on the brown dwarf CWW\,89Ab transiting a 3\,Gyr old Sun-like star. The team used the Spitzer/IRAC to observe the secondary eclipses, and they discovered a significant $9.3$-$\sigma$ overluminosity of CWW\,89Ab compared to the predicted values from the evolutionary models. The authors discussed the potential causes of this overluminosity, ruling out both stellar irradiation and tidal heating as likely explanations. They suggested that a dayside temperature inversion, resulting from a super-stellar C/O ratio, could be responsible for this phenomenon. If this is indeed the case, it would imply that CWW\,89Ab is a $36.5\,\mj$ object that formed via the core accretion process. The measurement of the C/O ratio would be essential to gain a better understanding of this system. It is also worth noting that, in contrast to BD-14\,3065b, CWW\,89Ab does not exhibit an inflated radius. We may also mention a recent study by \cite{Yan23} in which authors used ground-based high-resolution spectrograph CRIRES$^+$ to observe dayside thermal emission spectra of two ultra-hot Jupiters and detected strong CO emission lines in both planets. BD-14\,3065b is an ideal target for such a study.

The C/O ratio is an important characteristic proposed to constrain the formation of giant planets \citep{Oberg11}. Generally, a stellar C/O ratio is expected for companions that form through gravitational instabilities in a cloud, where all materials mix, and planets that form within the snow lines. A superstellar C/O ratio could indicate that the atmosphere formed from disk gas outside the water snowline, possibly via core accretion. It was later discussed that such simple division is not realistic and that several complex and interconnected processes affect the planetary formation, such as disc elemental composition, chemical evolution of the protoplanetary disc, planetary formation and migration, atmospheric evolution, and others (see \cite{Molliere22} for more details). It is still a long way off, if even possible at all, to connect atmospheric abundances with formation processes for individual systems. However, we can use the observed atmospheric composition to study the importance of various formation aspects in an attempt to reproduce observations. Furthermore, the atmospheric abundance constraints obtained with new and upcoming instruments will be crucial in developing planet formation models. Instruments such as GRAVITY, CRIRES$^+$, and recent and upcoming facilities such as JWST, ARIEL, and the ELT will obtain atmosphere composition for many planets.

BD-14\,3065b would be the most massive transiting companion with a measured C/O ratio to date, contributing to our understanding of giant planets and brown dwarfs. Furthermore, while WASP-18b, which orbits much closer to its star, has an atmosphere that is heated by stellar irradiation, resulting in an inverted temperature-pressure profile, BD-14\,3065b is expected to have an additional internal source. Hence, studying the temperature-pressure profile of its atmosphere can assist in understanding the processes involved in atmosphere heating.

\section{Summary} \label{sec:summary}

We have presented an analysis of the substellar object BD-14\,3065b transiting a subgiant F-type star. BD-14\,3065b has an orbital period of P = 4.2889731 ± 0.0000052 days, and the estimated age of the system is about 2 Gyr, making BD-14\,3065b, with a mass of $M_p=12.37\pm0.92\,\mj$ and a radius of $R_p=1.926\pm0.094\,\rj$, an extremely inflated object for its mass and age. This study focused on the two most notable differences between BD-14\,3065b and other systems in the literature, attempting to link them to the observed radius. Firstly, BD-14\,3065b is located at the deuterium burning boundary, and secondly, its primary star is an evolved subgiant. The conclusion drawn is that the radius could be explained by the enhanced rate of deuterium burning caused by sufficient heating of BD-14\,3065b's interior. Finally, detection of the secondary eclipse with TESS photometry enables a precise determination of the eccentricity and reveals BD-14\,3065b has a brightness temperature of $3520 \pm 130$\,K. Hence, BD-14\,3065b is considered an excellent target for future atmospheric characterization through secondary eclipse observations. Such an investigation would provide valuable insights into the processes responsible for the planet's atmospheric heating, as well as its formation history, by comparing its C/O ratio to that of its host star. This would be of significant value, given that BD-14\,3065b sits at the giant planet/brown dwarf boundary.

%
%
\begin{acknowledgements}
J.\v{S}. would like to acknowledge the support from GACR grant 23-06384O.
P.K. is grateful for the support from grant LTT-20015.
L.V. acknowledges ANID projects Fondecyt n. 1211162, QUIMAL ASTRO20-0025, and BASAL FB210003.
R.B. acknowledges support from FONDECYT Project 11200751 and additional support from ANID -- Millennium  Science  Initiative -- ICN12\_009.
J.A.C. acknowledges financial support from the Spanish Agencia Estatal
de Investigaci\'on (AEI/10.13039/501100011033) of the Ministerio de
Ciencia e Innovaci\'on and the European Regional Development Fund ``A
way of making Europe'' through project PID2022-137241NB-C42.

Funding for the TESS mission is provided by NASA's Science Mission Directorate.

We acknowledge the use of public TESS data from pipelines at the TESS Science Office and at the TESS Science Processing Operations Center.

Resources supporting this work were provided by the NASA High-End Computing (HEC) Program through the NASA Advanced Supercomputing (NAS) Division at Ames Research Center for the production of the SPOC data products.

This paper includes data collected by the TESS mission that are publicly available from the Mikulski Archive for Space Telescopes (MAST).

The PLATOSpec team would like to thank observers Marek Skarka, Ji\v{r}\'{i} Srba, Zuzana Balk\'{o}v\'{a}, Petr \v{S}koda and Lud\v{e}k \v{R}ezba.

This publication was produced within the framework of institutional support for the development of the research organization of Masaryk University.

This research has made use of the Simbad and Vizier databases, operated at the centre de Donn\'ees Astronomiques de Strasbourg (CDS), and of NASA's Astrophysics Data System Bibliographic Services (ADS).

This work has made use of data from the European Space Agency (ESA) mission {\it Gaia} (\url{https://www.cosmos.esa.int/gaia}), processed by the {\it Gaia} Data Processing and Analysis Consortium (DPAC, \url{https://www.cosmos.esa.int/web/gaia/dpac/consortium}). Funding for the DPAC
has been provided by national institutions, in particular, the institutions participating in the {\it Gaia} Multilateral Agreement.
\end{acknowledgements}

\bibliographystyle{aa}
\bibliography{astro_citations}
\onecolumn
\appendix
\renewcommand\thefigure{\thesection.\arabic{figure}}    
\section{Additional material}

\begin{longtable}{lccc}
\caption{Multi-order relative radial velocities of BD-14\,3065 from TRES and Pucheros+.} \label{tab:long} \\

\hline \multicolumn{1}{c}{\textbf{Date (BJD)}} & \multicolumn{1}{c}{\textbf{RV (m/s)}} & \multicolumn{1}{c}{\textbf{$\sigma_{RV}$ (m/s)}} & 
\multicolumn{1}{c}{\textbf{Spectrograph}} \\ \hline 
\endfirsthead

\multicolumn{4}{c}%
{{\bfseries \tablename\ \thetable{} -- continued from previous page}} \\
\hline \multicolumn{1}{c}{\textbf{Date (BJD)}} & \multicolumn{1}{c}{\textbf{RV (m/s)}} & \multicolumn{1}{c}{\textbf{$\sigma_{RV}$ (m/s)}} & \multicolumn{1}{c}{\textbf{Spectrograph}} \\ \hline 
\endhead

\hline \multicolumn{4}{r}{{Continued on next page}} \\ 
\endfoot

\hline 
\endlastfoot

2459601.934527 & 2479 & 181 & TRES \\
2459605.955459 & 2283 & 110 & TRES \\
2459616.890452 & 4647 & 172 & TRES \\
2459617.956610 & 3305 & 186 & TRES \\
2459618.882581 & 2362 & 153 & TRES \\
2459619.825369 & 2410 & 134 & TRES \\
2459620.899334 & 4586 & 133 & TRES \\
2459621.881599 & 4350 & 111 & TRES \\
2459622.928761 & 1979 & 107 & TRES \\
2459623.911069 & 2463 & 105 & TRES \\
2459624.901062 & 4245 & 112 & TRES \\
2459625.944000 & 4449 & 115 & TRES \\
2459626.800265 & 2637 & 120 & TRES \\
2459635.850469 & 2410 & 153 & TRES \\
2459922.954247 & -1124 & 183 & TRES \\
2459923.986662 & -2355 & 202 & TRES \\
2459924.995943 & 107 & 170 & TRES \\
2459930.979474 & -443 & 85 & TRES \\
2459932.988929 & -2176 & 81 & TRES \\
2459933.984636 & -304 & 135 & TRES \\
2459934.972431 & -249 & 140 & TRES \\
2459936.000038 & -2416 & 90 & TRES \\
2459936.965439 & -2776 & 123 & TRES \\
2459944.924846 & -2447 & 156 & TRES \\
2459948.961852 & -2354 & 101 & TRES \\
2459957.965266 & -2555 & 147 & TRES \\
2459963.949222 & -3 & 190 & TRES \\
2460013.776225 & -2499 & 173 & TRES \\
2460014.767013 & -1310 & 150 & TRES \\
2460016.749179 & -822 & 107 & TRES \\
2460017.788237 & -2601 & 88 & TRES \\
2460028.772600 & -849 & 170 & TRES \\
2460029.856313 & -1329 & 120 & TRES \\
2460030.758805 & -2344 & 123 & TRES \\
2460071.660621 & -508 & 153 & TRES \\
2460072.653619 & -998 & 133 & TRES \\
2460073.691053 & -1934 & 138 & TRES \\
2460074.681585 & -1450 & 114 & TRES \\
2460075.670057 & -298 & 105 & TRES \\
2460076.660340 & -286 & 120 & TRES \\
2460077.697400 & -1682 & 175 & TRES \\
2460252.019841 & 1010 & 119 & TRES \\
2460253.019642 & -142 & 93 & TRES \\
2460254.017140 & -384 & 99 & TRES \\
2460257.021429 & 526 & 94 & TRES \\
2460286.011982 & 1395 & 95 & TRES \\
2460287.012965 & 919 & 97 & TRES \\
2460288.027571 & -144 & 82 & TRES \\
2460289.014212 & -3 & 101 & TRES \\
2460352.826089 & 196 & 119 & TRES \\
2460353.857046 & 1840 & 80 & TRES \\
2460354.832224 & 1887 & 101 & TRES \\
2460355.799223 & 965 & 78 & TRES \\
2460410.726938 & 2405 & 126 & TRES \\
2460411.733248 & 1269 & 69 & TRES \\
2460412.719297 & 998 & 124 & TRES \\
2460413.666464 & 1585 & 70 & TRES \\
2460077.486053 & 2324 & 1122 & Pucheros+ \\
2460078.557373 & 715 & 1114 & Pucheros+ \\
2460078.580683 & 2164 & 1455 & Pucheros+ \\
2460080.569803 & 3522 & 973 & Pucheros+ \\
2460080.590706 & 3201 & 1857 & Pucheros+ \\
2460085.523472 & 3111 & 867 & Pucheros+ \\
2460085.544838 & 4610 & 901 & Pucheros+ \\
2460086.569549 & 1530 & 2006 & Pucheros+ \\
2460086.590926 & 2332 & 2117 & Pucheros+ \\
2460088.507060 & 5548 & 1197 & Pucheros+ \\
2460088.528426 & 3275 & 1204 & Pucheros+ \\
2460089.519780 & 4380 & 1107 & Pucheros+ \\
2460094.491713 & 3257 & 794 & Pucheros+ \\
2460096.531829 & 3333 & 374 & Pucheros+ \\
2460102.509028 & 4203 & 551 & Pucheros+ \\
2460287.740860 & 3158 & 511 & Pucheros+ \\
2460287.761727 & 4154 & 381 & Pucheros+ \\
2460326.827507 & 3543 & 508 & Pucheros+ \\
2460326.848822 & 2592 & 438 & Pucheros+ \\
2460327.816635 & 5211 & 444 & Pucheros+ \\
2460327.837962 & 5449 & 445 & Pucheros+ \\
2460328.818849 & 6205 & 577 & Pucheros+ \\
2460328.840164 & 6006 & 517 & Pucheros+ \\
2460358.833687 & 6854 & 355 & Pucheros+ \\
2460358.855014 & 6856 & 420 & Pucheros+ \\
2460359.831005 & 6576 & 566 & Pucheros+ \\
2460359.852322 & 6339 & 589 & Pucheros+ \\
2460364.842869 & 3092 & 512 & Pucheros+ \\
2460366.854662 & 6869 & 314 & Pucheros+ \\
2460367.841923 & 6927 & 502 & Pucheros+ \\
2460371.818079 & 7824 & 343 & Pucheros+ \\
2460372.743021 & 7685 & 494 & Pucheros+ \\
2460373.731877 & 3761 & 316 & Pucheros+ \\
2460374.639898 & 4474 & 243 & Pucheros+ \\
\hline

\end{longtable}

\begin{figure*}[ht!]
\centering
\includegraphics[width=1.0\textwidth]{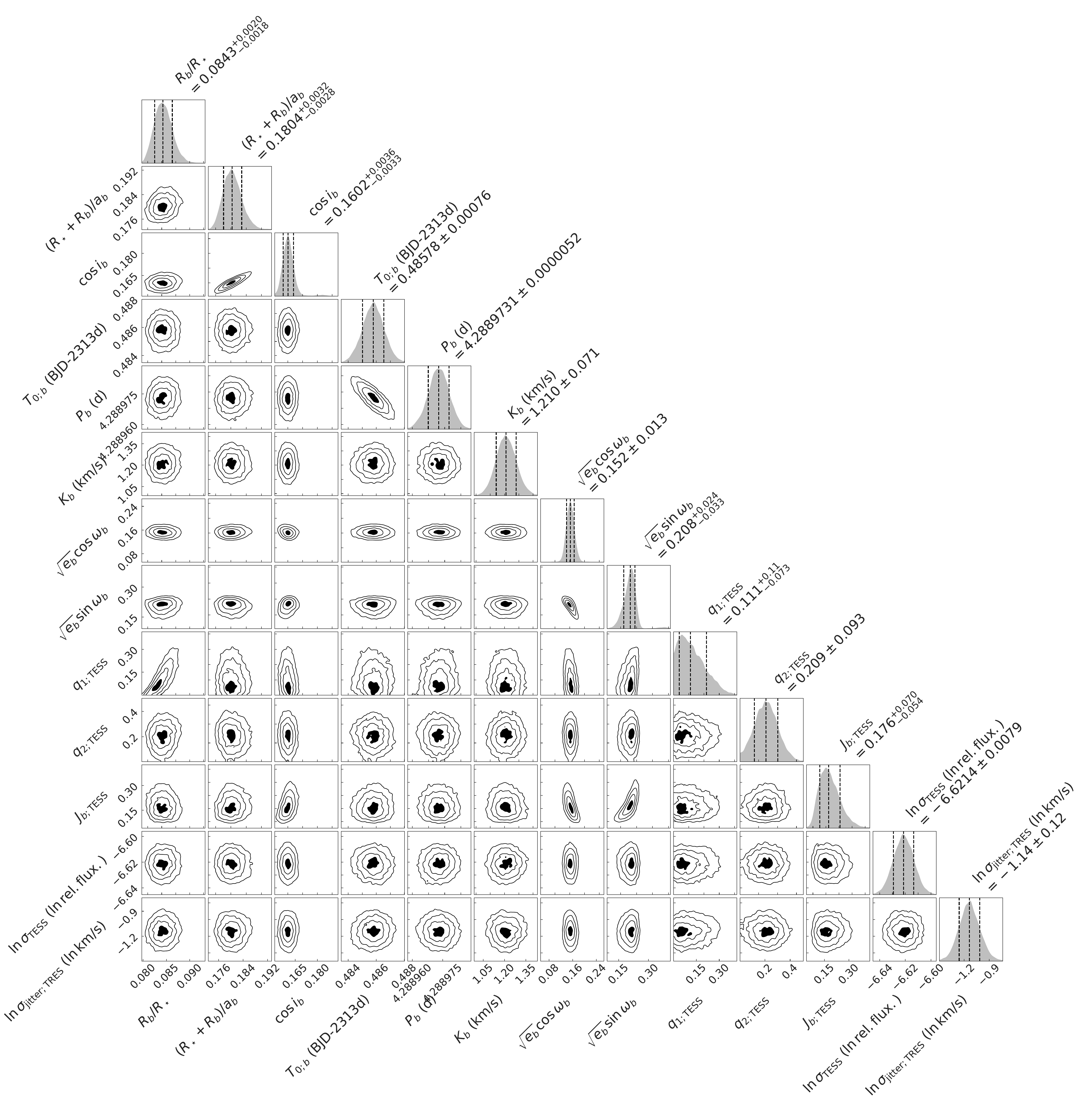}
\caption{The correlations between the free parameters of the 2-star model from the Nested Sampling. At the end of each row is shown the derived posterior probability distribution.} \label{fig:corner0}
\end{figure*}

\begin{figure*}[ht!]
\centering
\includegraphics[width=1.0\textwidth]{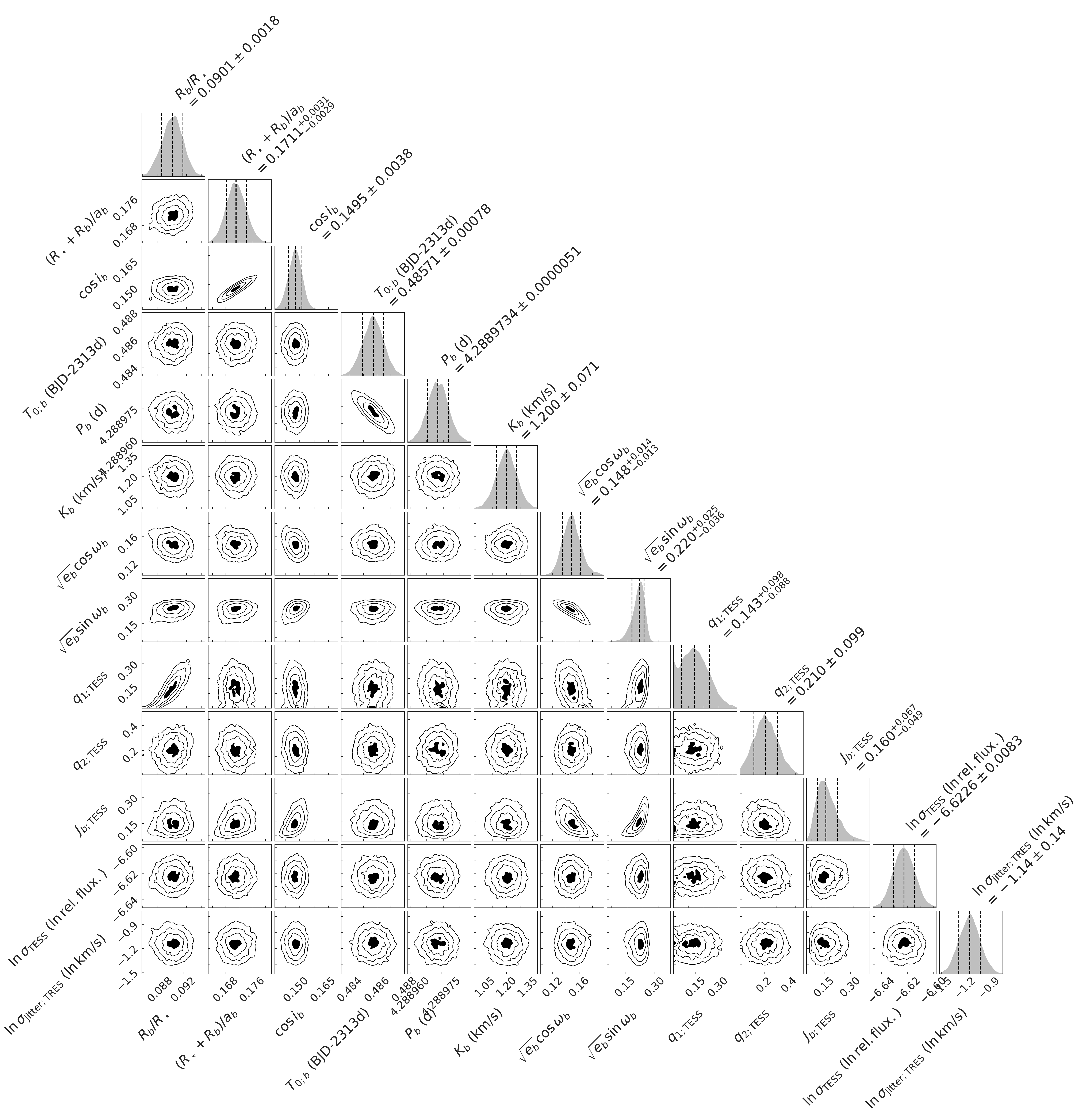}
\caption{The correlations between the free parameters of the 3-star model from the Nested Sampling. At the end of each row is shown the derived posterior probability distribution.} \label{fig:corner1}
\end{figure*}

\begin{figure*}[ht!]
\centering
\includegraphics[width=0.85\textwidth]{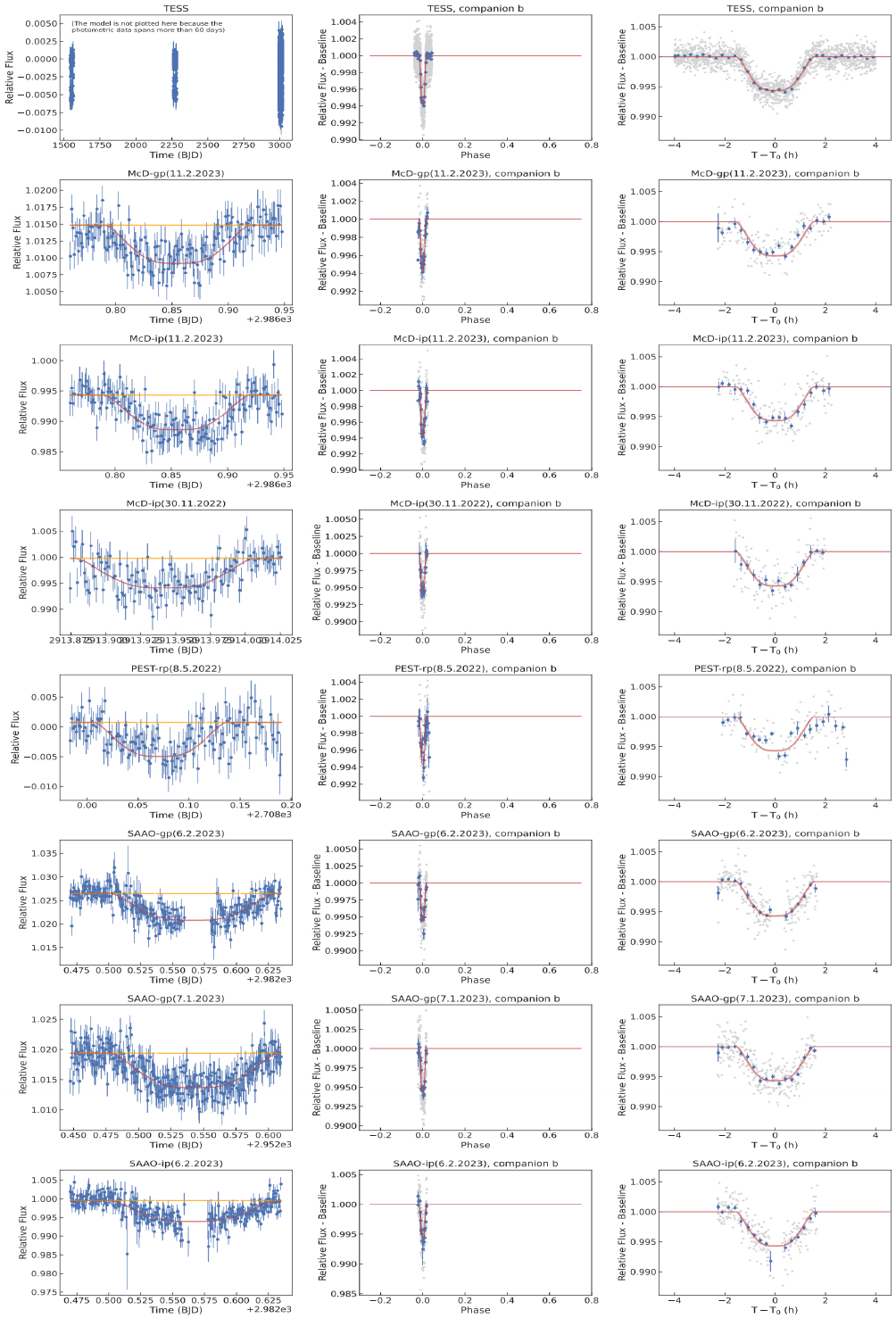}
\caption{The TESS and ground-based photometry fitted with the {\tt allesfitter}. The red line represents the best transit model, which is the same for each dataset.} \label{fig:phot_all}
\end{figure*}

\end{document}